\documentclass[conference]{IEEEtran}
\IEEEoverridecommandlockouts
\usepackage{cite}
\usepackage{amsmath,amssymb,amsfonts}
\usepackage{algorithmic}
\usepackage{pifont}% http://ctan.org/pkg/pifont
\usepackage{float}
\usepackage{subcaption}
\usepackage{array}
\usepackage[section]{placeins}
\usepackage{graphicx}
\usepackage{textcomp}
\usepackage{multirow}
\usepackage[justification=centering]{caption}
\usepackage[table]{xcolor}% http://ctan.org/pkg/xcolor
\usepackage{hyperref}
\usepackage{flushend}
\usepackage{tikz}
\usetikzlibrary{shapes,snakes}
\newlength{\leftLength}

\def\BibTeX{{\rm B\kern-.05em{\sc i\kern-.025em b}\kern-.08em
    T\kern-.1667em\lower.7ex\hbox{E}\kern-.125emX}}

\definecolor{lik_green}{HTML}{cbedc8}    
\definecolor{lik_blue} {HTML}{b4cfe4}
\definecolor{lik_red}  {HTML}{ffb7b3}

\usepackage{soul}
\usepackage{xargs}                      % Use more than one optional parameter in a new commands
\usepackage[colorinlistoftodos,prependcaption,textsize=tiny]{todonotes}
\newcommandx{\diego}[1] {\todo[linecolor=green, backgroundcolor=green!25, bordercolor=green] {\textbf{Diego:  }#1}}
\newcommandx{\daniel}[1]{\todo[linecolor=red,   backgroundcolor=red!25,   bordercolor=red]   {\textbf{Daniel: }#1}}

\newcommand{\summary}[2]{
\setlength{\leftLength}{0.5cm + (\widthof{{\small \textbf{#1}}}/2)}
\vspace{2ex}\noindent\begin{tikzpicture}
\node[align=center,draw,thin,minimum width=\columnwidth,inner sep=2.2mm] (titlebox)%
{\parbox{0.95\columnwidth}{\vspace*{0.25ex}\noindent\emph{#2}}};\node[fill=white] (W) at ([xshift=\the\leftLength] titlebox.north west) {{\small \textbf{#1}}};%
\end{tikzpicture}\vspace{2ex}}

\usepackage{hyperref}
\usepackage[firstpage]{draftwatermark}
 
\SetWatermarkAngle{0}
\SetWatermarkFontSize{8pt}
\SetWatermarkText{\raisebox{\textheight}{%
\begin{minipage}{\textwidth}%
\vspace{-18pt}\hfill%
\href{https://doi.org/10.5281/zenodo.5109401}{
\includegraphics[width=0.6in, height=0.6in]{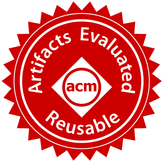}% IF REUSABLE BADGE GRANTED 
\includegraphics[width=0.6in, height=0.6in]{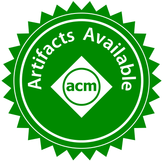}% IF AVAILABLE BADGE GRANTED 
}
\end{minipage}% 
}}

\begin{document}

\title{
Quality Guidelines for Research \\ 
Artifacts in Model-Driven Engineering
}

\author{\IEEEauthorblockN{Carlos Diego Nascimento Damasceno and Daniel Strüber}
\IEEEauthorblockA{\textit{Institute for Computing and Information Sciences} \\
\textit{Radboud University, Nijmegen, Netherlands}\\
\{d.damasceno,d.strueber\}@cs.ru.nl}
}

\maketitle

\begin{abstract}
Sharing research artifacts is known to help people to build upon existing knowledge, adopt novel contributions in practice, and increase the chances of papers receiving attention. In Model-Driven Engineering (MDE), openly providing research artifacts plays a key role, even more so as the community targets a broader use of AI techniques, which can only become feasible if large open datasets and confidence measures for their quality are available. However, the current lack of common discipline-specific guidelines for research data sharing opens the opportunity for misunderstandings about the true potential of research artifacts and subjective expectations regarding artifact quality. To address this issue,  we introduce a set of guidelines for artifact sharing specifically tailored to MDE research. To design this guidelines set, we systematically analyzed general-purpose artifact sharing practices of major computer science venues and tailored them to the MDE domain. Subsequently, we conducted an online survey with 90 researchers and practitioners with expertise in MDE. We investigated our participants' experiences in developing and sharing artifacts in MDE research and the challenges encountered while doing so. We then asked them to prioritize each of our guidelines as essential, desirable, or unnecessary. Finally, we asked them to evaluate our guidelines with respect to clarity, completeness, and relevance. In each of these dimensions, our guidelines were assessed positively by more than 92\% of the participants. To foster the reproducibility and reusability of our results, we make the full set of generated artifacts available in an open repository at \texttt{\url{https://mdeartifacts.github.io/}}.
\end{abstract}


\section{Introduction}

The term ``reproducibility crisis'' has gained traction as researchers from various fields have reported challenges to reproduce studies \cite{pashler_editors_2012}. 
Software engineering (SE) research is by no means exempt from this phenomenon, as many authors in SE have faced difficulties when trying to reproduce studies \cite{collberg_repeatability_2016}, performing replications with the direct support of authors \cite{lung_difficulty_2008}; and reusing artifacts in their own benchmarks \cite{glanz_codematch_2017}.
To mitigate these issues, SE conferences have started to incorporate artifact evaluation processes \cite{esecfse_call_2011,esecfse_esecfse_2020,icse_icse_2020,splc_call_2020,models_2020}.
In parallel to this, the Association for Computing Machinery has recently launched the \textit{ACM SIGSOFT Empirical Standards} document to communicate frequent expectations for research methods commonly used by their community \cite{ralph_acm_2020}. 

Software artifacts are known to provide means to others to build upon existing knowledge \cite{basili_building_1999}, adopt novel contributions in practice \cite{von_nostitz-wallwitz_towards_2018}, and increase the chances of papers receiving attention \cite{colavizza_citation_2020}. 
However, the current lack of discipline-specific guidelines for research data management \cite{marjan_grootveld_openaire_2018} opens the opportunity for conflicting subjective expectations toward artifact quality and hence, misunderstandings on the true potential of research artifacts \cite{hermann_community_2020}.

In Model-Driven Engineering (MDE), openly providing research artifacts plays a key role for the following reasons. 
First, despite some attempts to provide sets of models in consolidated repositories \cite{france_models_2006,babur_labeled_2019} and collections of UML models \cite{robles2017extensive,karasneh2016online} and transformations \cite{ATLzoo,struber2016scalability}, there is still a lack of large datasets of models of diverse modeling languages and application domains \cite{basciani2015model}. Having more systematic artifact sharing practices would help to increase the potential reuse of available models and support the evaluation of research tools and techniques.
Second, the need for consolidated artifact sharing practices in MDE research has recently become more pronounced, as the community targets a broader use of artificial intelligence (AI) techniques. Thus, to benefit from the advances in machine learning and even more so deep learning, MDE researchers should address the need for large open datasets and confidence measures for their quality.

In this paper, we present a set of guidelines for artifact sharing specifically tailored to MDE research. We designed and applied a comprehensive methodology to inform the design of our guidelines. This methodology included an analysis of available discipline-independent practices for artifact sharing from various SE venues, a study of literature on artifacts in the MDE domain, an application of project management principles from the literature, and an online survey with 90 practitioners and researchers with experience in MDE.
We address the following research questions:

\begin{description}
    \item [\textbf{RQ1:   }]    How can one define domain-specific guidelines for artifact sharing in the MDE domain?
    \item [\textbf{RQ2:   }]    How do the defined guidelines address the main challenges encountered by MDE experts?
    % \begin{itemize}
    %     \item [\textbf{RQ2.1: }]    What are the main challenges encountered?
    %     \item [\textbf{RQ2.2: }]    To what extent do the guidelines cover them?
    % \end{itemize}
    \item [\textbf{RQ3:   }]    How do MDE experts prioritize our practices?
    \item [\textbf{RQ4:   }]    What is the quality of the proposed guidelines?
    % \begin{itemize}
    %     \item [\textbf{RQ4.1: }]    How complete are the guidelines?
    %     \item [\textbf{RQ4.2: }]    How clear are the guidelines?
    %     \item [\textbf{RQ4.2: }]    How relevant are the guidelines?
    % \end{itemize}
\end{description}

The outcome is a set of 84 practices, structured along seven factual questions (i.e., what, why, where, who, when, how, how much/many), and prioritized into the three levels \textit{essential}, \textit{desirable}, and \textit{unnecessary}. The quality of our guidelines in terms of completeness, clarity, and relevance was assessed positively by more than 92\% of our participants. These findings indicate that our guidelines can be useful for guiding authors in preparing high-quality research artifacts. Furthermore, we believe that our guidelines can inform organizers of artifact evaluation processes of MDE conferences and journals on questions to keep in mind when reviewing artifacts and on practices generally accepted in the community.

This paper is organized as follows:
In Sect.~\ref{sec:bg}, we discuss the background concepts that underpinned our study. 
In Sect.~\ref{sec:method}, we present the methodology used to address our investigations.
In Sect.~\ref{sec:results}, we report the results of our survey and participants' responses.
In Sect.~\ref{sec:discussion}, we discuss the implications of our findings to artifact authors and reviewers, opportunities for improvement, and threats to validity.
In Sect.~\ref{sec:relwork}, we enumerate some related work.
In Sect.~\ref{sec:conclusion}, we close this paper and discuss future work.

\section{Background}
\label{sec:bg}

\subsection{Research Data Management and Artifact Sharing}

As Computer Science (CS) becomes more important in all branches of the modern sciences,
software artifacts are now seen as another important outcome of scientific research \cite{hermann_community_2020}.
They are necessary evidence to validate findings and results in most of modern sciences \cite{wilson_good_2017}, help researchers building upon existing knowledge \cite{basili_building_1999}, encourage the adoption of novel research in practice \cite{von_nostitz-wallwitz_towards_2018}, and increase the chances of citations \cite{colavizza_citation_2020}. 
However, 
As numerous SE researchers report increasing challenges to reproduce studies \cite{lung_difficulty_2008,collberg_repeatability_2016,glanz_codematch_2017}, the ``reproducibility crisis'' known in natural sciences and humanities \cite{pashler_editors_2012} is now an imminent threat to the reproducibility, replicability, and repeatability of empirical experimentations in SE \cite{wohlin_experimentation_2012}.

Research Data Management (RDM) helps investigators to make conscious decisions about research data \cite{ru_research_2021}. It comprises various activities, from data planning and synchronization, over elements standardization, to data sharing and repository management; which encourage research reproducibility and software sustainability \cite{van_eeuwijk_research_2021}.
Funding agencies and research councils have started to incorporate requirements for data management and sharing as a standard part of their policies. Such initiatives can benefit research environments to promote their scholarly work, enable new collaborations, allow maximum use of contributed information, and advance science to the benefit of the whole society \cite{corti_managing_2019}. 
However, due to the multi-faceted nature of software artifacts in terms of purpose, size, complexity and format, RDM policies may still need to be adapted \cite{marjan_grootveld_openaire_2018} to provide clear, complete and relevant domain-specific guidance.

\subsection{Artifacts in Model-Driven Engineering}

In Model-Driven Engineering, rigorous abstractions models are used as central artifacts in systems' design and development process \cite{brambilla_modeldriven_2012}. Typical artifacts in MDE include models, metamodels, model transformations, and modeling tools \cite{basso_revisiting_2017}. As such distinct assets, MDE artifacts should comply with their requirements and quality criteria, such as model semantics and syntax \cite{krogstie_quality_2012} and technology standards \cite{steinberg_emf_book_2008}.
Basso et al. \cite{basso_revisiting_2017} suggest that MDE artifacts should report information about aspects such as their context of use, usage restrictions, and business opportunities. On top of technical information, organizational and social factors should also be considered as sources 
of issues that may affect the adoption of MDE tools and artifacts \cite{whittle_taxonomy_2017}, such as {sustainability over the long term}, {appropriateness for repurposing}, and {opportunities of interacting with authors/community}. Studies like these constitute useful resources to guide MDE researchers and practitioners on creating and sharing high-quality research datasets, even more so as the MDE community aims at developing effective and efficient AI-based MDE techniques, a task that can only become feasible if large, open, and good-enough benchmark sets are available.

\subsection{Project Quality Management}

In a survey of more than 1.5k PhD students from Europe and Israel, the majority of the doctoral students reported having no training or expertise in managing research projects \cite{katz_challenges_2016}. As doctoral research projects meet the same formal definition of a project \cite{pmi_guide_2017}, researchers suggest that PhD students should receive basic project management lessons to better manage their research as they move towards the successful completion of their doctorate \cite{katz_challenges_2016}.

According to the Project Management Institute (PMI), project management (PM) is defined as the application of knowledge, skills, tools, and techniques to project activities to meet the project requirements \cite{pmi_guide_2017}. Project quality management is a PM knowledge area that applies to all projects, regardless of the nature of their deliverables (i.e., artifacts). It aims at incorporating the organization’s quality policy regarding planning, managing, and controlling project and product quality requirements so that stakeholders’ expectations are met. Quality measures and techniques are specific to the type of artifact being produced and should be always identified and documented \textit{a priori}.

\textit{Plan quality management} is the process of identifying quality requirements and/or standards for a project and its artifacts and documenting means that a project shall demonstrate compliance with such quality requirements and/or standards. Failure to identify and meet quality requirements can have serious negative consequences for project stakeholders. To understand the concept of quality within the context of a specific project, there is an extensive toolbox of methods \cite{tague_quality_2005} from which the 5W2H method constitutes a simple but powerful framework for research planning, analysis, or reviewing.

\begin{figure*}[!t]
    \centering
    \includegraphics[width=\linewidth]{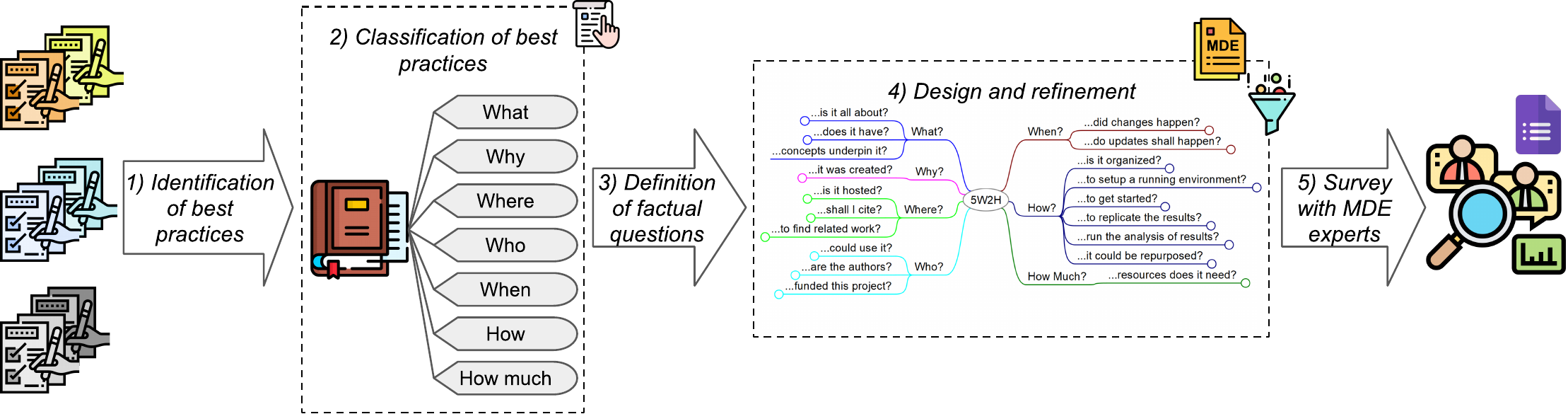}
    \caption{Designing a set of quality guidelines for MDE research artifact sharing}
    \label{fig:paper_schema}
\end{figure*}

The term ``5W2H'' is an abbreviation of seven keywords: \textbf{W}hat, \textbf{W}here, \textbf{W}hy, \textbf{W}ho, \textbf{W}hen, \textbf{H}ow, and \textbf{H}ow Much.
This is a well-known method by journalists to report news \cite{pan_framing_1993,hart_fiveWs_1996}.
From their perspective, reporters are expected to gather and present categories of information to their audience. These categories should indicate the essential information that people may want to know about a news story. In the literature, authors also refer to this method as the Five Ws \cite{hart_fiveWs_1996} or 5W1H \cite{pan_framing_1993}.

In project quality management \cite{pmi_guide_2017}, the 5W2H method can be used for asking questions about a process or problem. The 5W2H structure forces managers to consider all aspects of the situation when analyzing a process for improvement opportunities, a problem is identified but must be better defined, lanning a project or steps of a project, or reviewing a project after completion \cite{tague_quality_2005}.

According to Tague's book ``The Quality Toolbox'' \cite{tague_quality_2005}, the 5W2H method works as follows: 
(i.)   Review the situation under study and make sure the subject of the 5W2H is understood.
(ii.)  Develop appropriate factual questions about the situation for each keyword.
(iii.) Answer each question. If answers are not known, create a plan for finding them.
(iv.)  What you do next depends on your situation.
If you are planning a project, your factual questions and answers should help form your plan.
If you are analyzing a process for improvement opportunities, your questions and answers should lead to additional questions about possible facts. 
If you are reviewing a completed project, your factual questions and answers should lead to additional questions about modifying, expanding, or standardizing something.

\section{Methodology}
\label{sec:method}

To take full benefit from artifact sharing, research communities and institutions should join efforts to establish common standards for data management and publishing \cite{the_royal_society_openscience_2012}. However, this is a non-trivial task since the expectations toward artifacts vary depending on communities, roles, and artifact types \cite{hermann_community_2020}.

To design and evaluate a guideline set for quality management of MDE research artifacts, our methodology was informed by five major data sources:
(i.) existing general-purpose and discipline-independent quality guidelines of major venues in CS and SE;
(ii.) project management literature focusing on quality management \cite{pmi_guide_2017,tague_quality_2005} and the 5W2H method \cite{tague_quality_2005}; 
(iii.) MDE literature in tooling issues \cite{whittle_taxonomy_2017} and modeling artifact repositories \cite{basso_revisiting_2017};
(iv.) our own experiences---one author has 10 years of experience in the domain, including membership in artifact evaluation committees (AEC) for 3 MDE-related conferences;
(v.) an online survey with MDE experts.

Based on these data sources, we developed a systematic five-phase methodology, schematically depicted in Fig.~\ref{fig:paper_schema}.
Our methodology employed the following phases:
(i.) identification of practices from guidelines for artifact sharing,
(ii.) categorization of practices based on the 5W2H method for quality management \cite{tague_quality_2005},
(iii.) definition of factual questions to inquire about practices,
(iv.) design and refinement towards a domain-specific guidelines for MDE artifact quality management, and
(v.) evaluation and prioritization of the guidelines by MDE experts.
Details about each phase are provided in the next subsections.
To foster reproducibility and reusability, we made an online supplementary material available on Zenodo \cite{mdeartifacts_zenodo}, Github \cite{mdeartifacts_github}, and in our project website \cite{mdeartifacts_website}. 

\subsection{Identification of practices for artifact sharing}

To elicitate the quality expectations for research artifacts in the broader field, we analyzed eight guidelines sets for artifact sharing of major CS and SE publishers, venues, and organizations, namely:

\begin{enumerate}
    \item The ACM Artifact Review and Badging \cite{acm_artifact_2020}
    \item The EMSE Open Science Initiative \cite{mendez_fernandez_open_2019,monperrus_how_2019,emse_emse_2021,emse_emse_2021-1}
    \item The Journal of Open Science Software (JOSS) \cite{katz_publish_2018} 
    \item The Journal of Open Research Software (JORS) \cite{jors_journal_2021}
    \item The Guidelines by Wilson et al. \cite{wilson_good_2017}
    \item The NASA Open Source Software Projects \cite{nasa_nasa_2021}
    \item The TACAS artifact evaluation guideline  \cite{tacas_tacas_2019}
    \item The CAV artifact evaluation guideline \cite{cav_artifacts_2019}
\end{enumerate}

\begin{figure}[!ht]
    \centering
    \includegraphics[width=\linewidth]{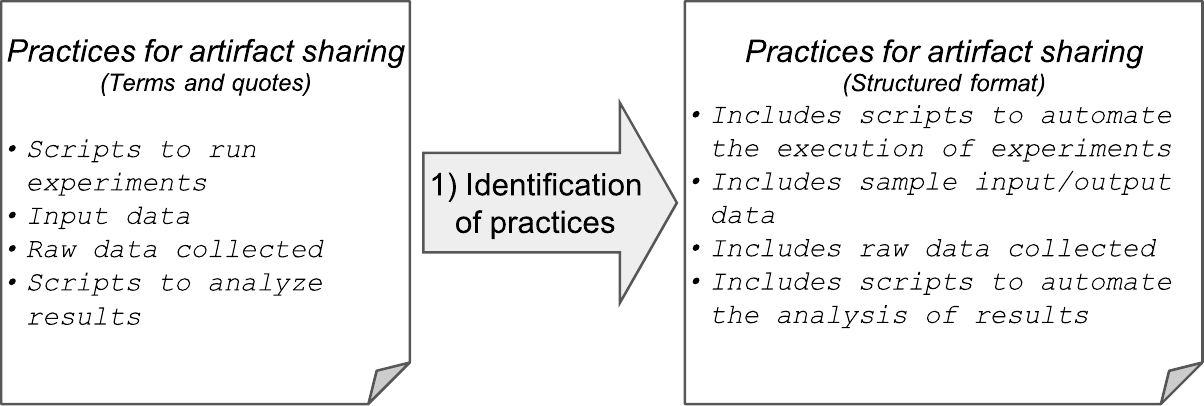}
    \caption{Examples of extracted and refined practices}
    \label{fig:step1_restruct}
\end{figure}

From these guidelines sets, we obtained an initial list of 284 general-purpose practices.
This list was compiled in two steps. First, we analyzed each guideline set to extract practices and recommendations for artifact sharing. Second, we refined these extracted practices by standardizing their structure and terminology. In Fig.~\ref{fig:step1_restruct}, we show a sample of the practices extracted and their refined versions.

\subsection{Categorization of best practices according to the 5W2H}

To address the multitude of expectations for artifacts' usage and quality criteria, we employed the 5W2H method \cite{tague_quality_2005} to categorize the extracted practices using the five Ws and two Hs as content tags. 
% The 5W2H is a well-known tool used by journalists, researchers and project managers to report and structure information according to seven perspectives.
% 
We take research practices as \textit{answers to factual questions} that 
researchers or reviewers could ask about an artifact. 

We adapted the 5W2H framework to the context of research artifacts and formulated a pattern to label research practices according to its perspectives. 
In Table~\ref{tab:5w2h_artifact}, we show examples of categorized practices and their respective labels.

\begin{table}[!ht]
\centering
\caption{Examples of practices and their 5W2H labels}
\resizebox{0.5\textwidth}{!}
{% <------ Don't forget this %
\begin{tabular}{r|c} % >{\centering\arraybackslash}p{0.85\linewidth}|
\hline
\textbf{Label}& \textbf{Best practice} \\ \hline
What          & Provides an indication of the context of the software use\\ \hline
Where         & Provide info on how to cite the project (e.g., CITATION file)\\ \hline
When          & Provides explanation for changes (e.g., CHANGELOG, commit)\\ \hline
Who           & Uses open/non-proprietary file formats\\ \hline
How           & There are scripts for every stage of data processing\\ \hline 
How           & Provides suggestions for other potential applications\\ \hline
How Much      & Provides a way to replicate the results with modest resources\\ \hline
\end{tabular}
}
\label{tab:5w2h_artifact}
\end{table}

The main goal of this step was to gain insights on the types of factual questions that the extracted practices could possibly address. In this task, we employed mind mapping \cite{tague_quality_2005} to structure our practices into branches labeled with one of the 5W2H perspectives. We found the three major categories, namely \textit{How}, \textit{What}, and \textit{Where}, composed 83.28\% of our initial set of practices. In Table~\ref{tab:5w2h_total}, we show the percentage of practices categorized in each perspective.

\begin{table}[!ht]
    \centering
    \caption{Percentage of practices in each 5W2H perspective}
    \begin{tabular}{r|r}
    \hline
    Perspective &   \%  \\ \hline
    What        & 31.44 \\ \hline
    Where       & 17.00 \\ \hline
    Why         &  3.97 \\ \hline
    Who         &  7.93 \\ \hline
    When        &  2.83 \\ \hline
    How         & 34.84 \\ \hline
    How Much    &  1.98 \\ \hline
    \end{tabular}
    \label{tab:5w2h_total}
\end{table}

In this labeling process, we used the following pattern:
As part of the \textit{What} perspective, we assigned practices associated with the overall description, context and content of the artifact.
As part of the \textit{Where} perspective, we assigned practices associated with repository hosting, artifact citation and related work.
As part of the \textit{Why} perspective, we assigned practices associated with the reasoning to create an artifact, its objectives and main advantages.
As part of the \textit{Who} perspective, we assigned practices associated with usage rights, licensing, authors' details, and funding agencies.
As part of the \textit{When} perspective, we assigned practices associated with version control and identification, updates, and future plans.
As part of the \textit{How} perspective, we assigned practices associated with the environment setup, replications, analysis of results, and repurposing.
Finally, as part of the \textit{How much} perspective, we assigned practices associated with quantitative information about system requirements and the time needed to run the artifact.

\subsection{Definition of intermediate factual questions}

After labeling our practices, we used their associated 5W2H tags to elaborate factual questions 
that researchers and reviewers could eventually ask about a given research artifact, e.g., 
``\textbf{What} is it about?'', ``\textbf{Where} shall I cite?'', ``\textbf{Who} are the authors?''.
These questions were designed to help researchers to systematically think about concerns in artifact sharing and provide directions to additional improvement questions.
The mind map in Fig.~\ref{fig:paper_schema} illustrates our factual questions and their associated perspectives. As we discuss in the next section, these questions were designed to kick off the creation of our domain-specific guidelines.

\subsection{Design and refinement of the MDE-specific guidelines}

Getting artifacts into publishable shape is often perceived as difficult and time-consuming \cite{timperley_understanding_2020}. To cope with time and resource constraints in research projects, we reviewed our guidelines to identify, match and merge similar practices. After analyzing our initial set of 284 practices, we found various redundant items that were common to different catalogs or covered related issues. Based on these similarities, we elaborated 77 practices to cover one or more items from our initial set of recommendations. 

At this point, our guidelines included practices addressing concerns for general types of artifacts, e.g. version control, user instructions, that also apply to MDE artifacts. However, MDE-specific aspects which are known to be important, e.g., model semantics, syntax, and technologies; and influence on the quality of models \cite{krogstie_quality_2012}, were still missing.

To tailor our guidelines to the MDE domain, we relied on our own experiences and analyzed two studies from the MDE literature: a taxonomy of tool-related issues affecting the adoption of MDE in the industry \cite{whittle_taxonomy_2017}, and a study on quality criteria for repositories of modeling artifacts \cite{basso_revisiting_2017}. Based on these two studies, we elaborated seven extra practices covering MDE specific concerns and assigned them to an additional factual question inquiring \textit{``What concepts and technologies underpin the artifact?''}. Finally, we incorporated two factual questions which covered associated tasks: ``\textit{How to compile/build?}'' and \textit{``How to setup a running environment?''}.

\begin{figure*}[!ht]
    \centering
    \includegraphics[width=0.7\linewidth]{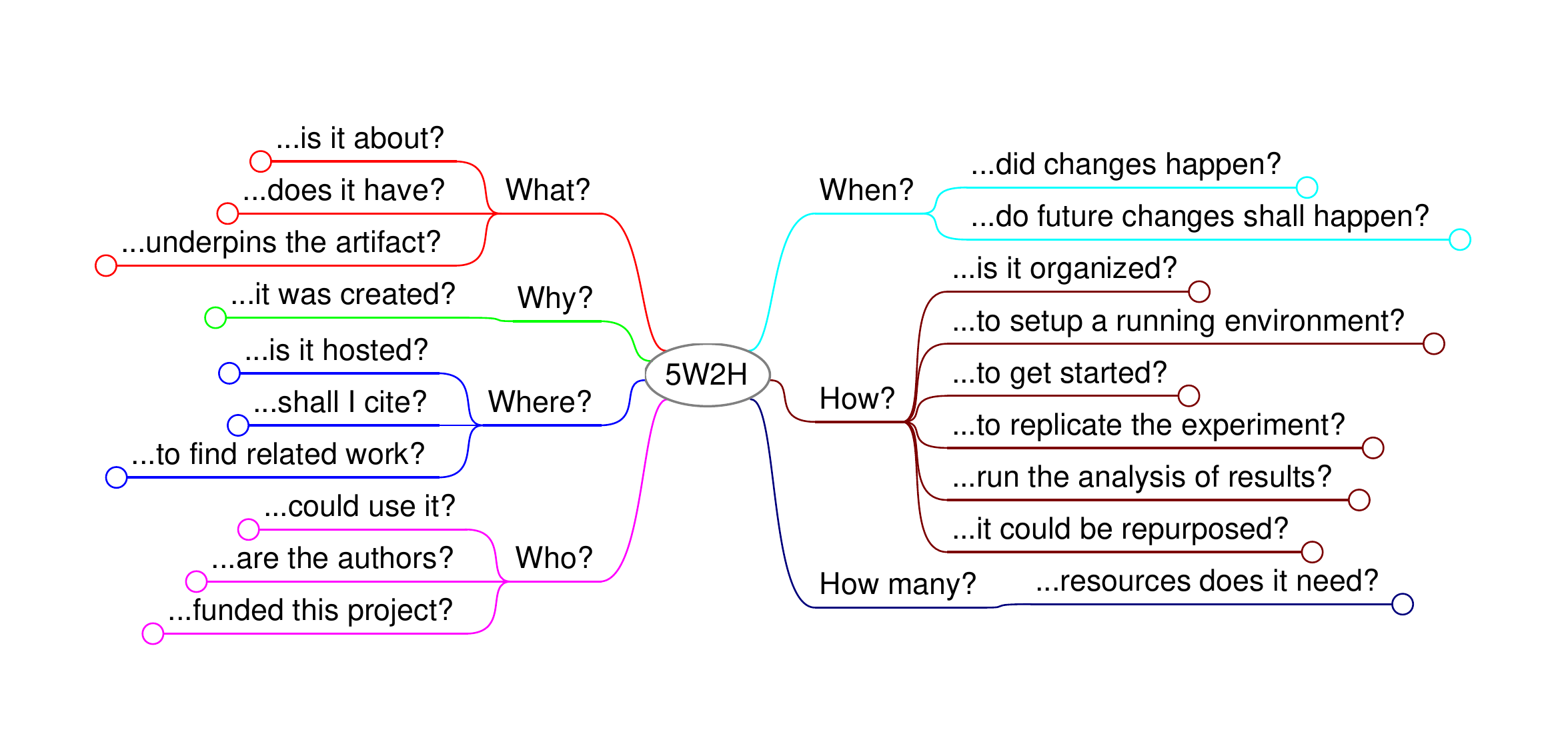}
    \caption{Mind map with the final set of 19 factual questions}
    \label{fig:step4_5w2h}
\end{figure*}

This led to our final set of guidelines with 19 factual questions (shown in Fig.~\ref{fig:step4_5w2h}) and 84 practices. The full set of practices and questions is available in our website \cite{mdeartifacts_website}.
We provide traceability from the 284 analyzed to our 84 practices in our associated artifact (\textit{practices4mde\_03Final.pdf} in \cite{mdeartifacts_zenodo}).

\subsection{Survey}

Developing useful research artifacts is challenging as people may have different expectations depending on their role and experience \cite{hermann_community_2020}. Thus, we designed a questionnaire survey for MDE experts to study their challenges encountered (RQ2) and to ask them to assess and prioritize our guidelines (RQ3--4). In Table~\ref{tab:survey_structure}, we show an overview of our survey.

\begin{table}[!ht]
    \centering
    \caption{Overview of the survey}
    \resizebox{0.475\textwidth}{!}
    {% <------ Don't forget this %
        \begin{tabular}{>{\centering\arraybackslash}p{0.20\linewidth}|p{0.75\linewidth}}
        \hline
        Topic& Description\\\hline
        {Demographics data}&
        Questions about the participants (Q1) gender and their (Q2) current primary role
        \\\hline
        {General experiences with artifacts}&
        How would you rate your experience in (Q3) artifact development and sharing and (Q4) reusing artifacts in MDE research?;
        (Q5) Have you ever submitted an artifact for evaluation?
        Have you ever 
        (Q6) contacted other researchers or 
        (Q7) been contacted by other researchers 
        asking for help on reusing their artifacts?
        \\\hline
            {Challenges in artifact sharing}&
        (Q8) Which challenges have you encountered during the sharing and use of artifacts in MDE research projects? 
        \\\hline
        {Evaluation of the Guidelines}&
        We asked participants to rate the (Q9-34) relevance of each one of the 84 practices and,
        if needed, recommend additional guidelines.
        \\\hline
        {Final evaluation}&
        How do you assess the (Q35) clarity, (Q36) completeness, and (Q37) relevance of these guidelines?
        Open field for (Q38) additional remarks or (Q39) providing e-mail, if wanted to stay updated about our results.
        \\\hline
        \end{tabular}
    }
    \label{tab:survey_structure}
\end{table}

\smallskip
\noindent{}\textbf{Participant recruitment.} 
We performed our survey in April-May 2021. 
Participants were recruited in two main ways:
We invited 335 people via e-mail and distributed the invitation on relevant online channels. The majority of personally invited participants were chosen for having taken a part in the MODELS AEC, coauthored a MODELS paper that earned an ACM artifact badge, and/or coauthored a Software and Systems Modeling (SoSyM) journal paper including some artifact in the last three years. In addition, we invited personal contacts from the MDE community, and encouraged our invitees to forward the invitation to their own contacts with relevant MDE experience. Online channels on which we distributed the call were the PlanetMDE mailing list \cite{planetmde_planetmde_2021}, and our personal LinkedIn and Twitter accounts. Our recruitment activities led to a total of 90 responses.

\smallskip
\noindent{}\textbf{Questionnaire design.} 
We designed 39 questions to understand the participants' background and to evaluate the clarity, completeness, and relevance of our guidelines. The questions in our survey covered five topics, as shown in Table~\ref{tab:survey_structure}.

First, we collected demographic information, specifically, their gender and current primary role. Demographics are useful for understanding the context of survey's participants.

Second, we inquired our participants about their general artifact sharing experiences. We asked them to report their level of experience with research artifact sharing and reuse, both on a 5-points Likert scale. We asked if they have ever submitted an artifact for evaluation and if they have contacted another researcher to reuse artifacts. For the latter, we inquired if they have ever contacted a researcher while trying to reuse an artifact or have ever been contacted by some researcher asking for support with an artifact they previously published. These questions were designed to evaluate the collective experience of our participants with research artifacts.

Third, to gather a domain-specific understanding of what makes MDE artifact sharing and development challenging, we asked the participants to report on challenges encountered during these activities, using an open text field. This field was included to identify issues that could complement our understanding of artifact sharing in MDE research. Based on these responses, 
we aimed to address \textbf{RQ2} by drawing a picture of the challenges faced in MDE artifact sharing and analyzing to what extent our guidelines covered these issues.

Fourth, we asked our participants to prioritize and evaluate our guidelines. We presented all 84 practices to our participants as follows: using the 5W2H perspectives as main categories, each perspective was refined into several factual questions with associated practices proposed as means to address them. We asked our participants to rate each practice as either ``Essential'', ``Relevant'', or ``Unnecessary'', providing a ``no answer'' option for participants who did not want to rate the practice at hand. To capture any factual questions and practices missed, we also provided an open text field to collect suggestions. These questions were designed to categorize our practices according to their priority and hence, address  \textbf{RQ3}.

Fifth and finally, we asked our participants to provide an overall score to our guidelines. To this end, we first recapitulated the full set of 19 factual questions. We then asked to the participants to evaluate our guidelines considering dimensions: clarity, completeness, and relevance. For each dimension, to obtain a nuanced assessment, participants were asked to specify a score on a 7-point Likert scale. The scale end-points were labeled, in the case of clarity as ``very unclear (1)'' and ``very clear (7)'', and similarly for completeness and relevance.
To collect useful information for interpreting the given scores, we provided an open text field for additional remarks. With this set of scores and additional remarks, we addressed \textbf{RQ4}. For participants interested in receiving information about the results of our survey, we provided a text field asking for their e-mail addresses. To counter possible bias, we informed our participants that we would remove the e-mail addresses from our collected data before processing it.

We used the Google Forms platform to perform the survey. In our dry runs, completing the questionnaire took around 15 minutes, which we communicated as an estimate.

\section{Results}
\label{sec:results}

We now present our results: our practices and the insights about them brought forward in our MDE expert survey. The presentation of results is organized into {five} parts. First, we give a brief overview of our practices (addressing RQ1). Second, we discuss our respondents' demographic characteristics and experiences with research artifacts. Third, we analyze the challenges reported by our participants on artifact sharing and reuse, and how our guidelines address them (RQ2). Fourth, we analyze how our participants prioritized our set of practices (RQ3).
Finally, we present our participants' assessment of the completeness, clarity, and relevance of our guidelines (RQ4).

\subsection{Overview of our guidelines for MDE artifact sharing (RQ1)}

Our guidelines for artifact sharing in MDE research comprise a structured set of 84 best practices. These best practices are proposed as answers to 19 different factual questions (shown in Fig.~\ref{fig:step4_5w2h}) that researchers may ask about an MDE research artifact. These factual questions cover the seven perspectives of the 5W2H framework and should encourage researchers to inquire about artifact's quality concerns. Table~\ref{tab:5w2h_priority} shows a selection of our practices based on a prioritization from our survey participants (explained in Sect.~\ref{sec:rq3}).

While our guidelines aim to cover all relevant aspects of MDE artifact sharing in an encompassing manner, we acknowledge that they might need to be tailored to particular circumstances. For example, for an artifact that is not executable (such as a collection of models), some guidelines in category \textit{6.4) How to replicate the experiment?} may not be applicable. Users of the guidelines, such as artifact authors and artifact evaluation organizers, should reflect on the guidelines and apply them in a way that is meaningful in their particular circumstances.

\summary{Guidelines for MDE artifact sharing (RQ1)}{
We derived a set of 84 practices, structured along 19 factual questions, to provide guidance to artifact sharing in MDE research. Our guidelines are proposed as means to address factual questions that researchers may inquire about an MDE artifact and visualize ``artifact quality'' from different perspectives.
}

\subsection{Survey demographics and experiences with artifacts}

Based on our recruitment activities, we obtained a total of {90 responses}.
In Table~\ref{tab:demographics}, we capture an overall picture of our respondents' gender and current primary role. 

\begin{table}[!ht]
\centering
\caption{Demographics - Primary role and gender}
% \resizebox{0.45\textwidth}{!}
{% <------ Don't forget this %
% \def\arraystretch{1.20}%  1 is the default, change whatever you need
% \begin{tabular}{c|l|r|r}
% \hline
% Gender & Primary role            & \# & \% \\ \hline
% \multirow{3}{*}{Female} & Academic (Pre-Phd)      & 4  &  4.4     \\ 
% & Academic (Professor)    & 4  &  4.4     \\ 
% & Academic (Post-Doc)     & 5  &  5.6     \\ \hline
% \multirow{5}{*}{Male}   & Industrial Practitioner & 7  &  7.8     \\
% & Industrial Researcher   & 7  &  7.8     \\ 
% & Academic (Pre-Phd)      & 10 & 11.1     \\ 
% & Academic (Post-Doc)     & 18 & 20.0     \\ 
% & Academic (Professor)    & 35 & 38.9     \\ 
% \hline
% \end{tabular}
\begin{tabular}{c|c|c}
\hline
Primary role            & \# Male & \# Female \\ \hline
Industrial Practitioner & 7       & 0         \\ \hline
Industrial Researcher   & 7       & 0         \\ \hline
Academic (Pre-Phd)      & 10      & 4         \\ \hline
Academic (Post-Doc)     & 18      & 5         \\ \hline
Academic (Professor)    & 35      & 4         \\ \hline
\end{tabular}
}
\label{tab:demographics}
\end{table}

In our poll, 43.3\% identified themselves as \textit{Academic Professor} and 85.6\% were identified as male. While roles were more evenly distributed among \textit{Female academics}, the majority of our male respondents were academic professors. Although we provided an open text field for non-binary genders, no participant used it.

Regarding our participants' experiences with artifact reuse, 83.2\% reported having either made contact with another researcher or been contacted by another researcher asking for support with an artifact. These findings indicate that our participants have meaningful collective experiences   with research artifacts. In Table~\ref{tab:auth_contact}, we show the numbers of participants who have made contact with or been contacted by another researcher.

\begin{table}[!ht]
\centering
\caption{Have you made contact/contacted someone for the purpose of artifact reuse?}
% \resizebox{0.45\textwidth}{!}
{% <------ Don't forget this %
\begin{tabular}{c|c|r|r}
\hline
Made contact? & Been contacted?& \# & \%   \\ \hline
No         & Yes             & 13 & 14.4 \\ 
Yes        & No              & 13 & 14.4 \\ 
No         & No              & 15 & 16.7 \\ 
Yes        & Yes             & 49 & 54.4 \\ 
\hline
\end{tabular}
}
\label{tab:auth_contact}
\end{table}

\subsection{How do the proposed guidelines address challenges encountered by MDE experts? (RQ2)}

In this section, we discuss the top ten challenges reported by MDE experts and how our guidelines cover these issues. To understand how our guidelines address the challenges faced by MDE experts, we first analyzed the answers to an open-ended question we provided to identify ``\textit{issues that make the sharing and use of artifacts difficult}''. From our 90 participants, we obtained 66 answers that we analyzed, tagged, and compared against our guidelines. The full set of 66 answers is available as supplementary material \cite{mdeartifacts_zenodo,mdeartifacts_github}.

\subsubsection{What are the challenges encountered?}

\begin{figure}[!ht]
    \centering
    \includegraphics[width=\linewidth]{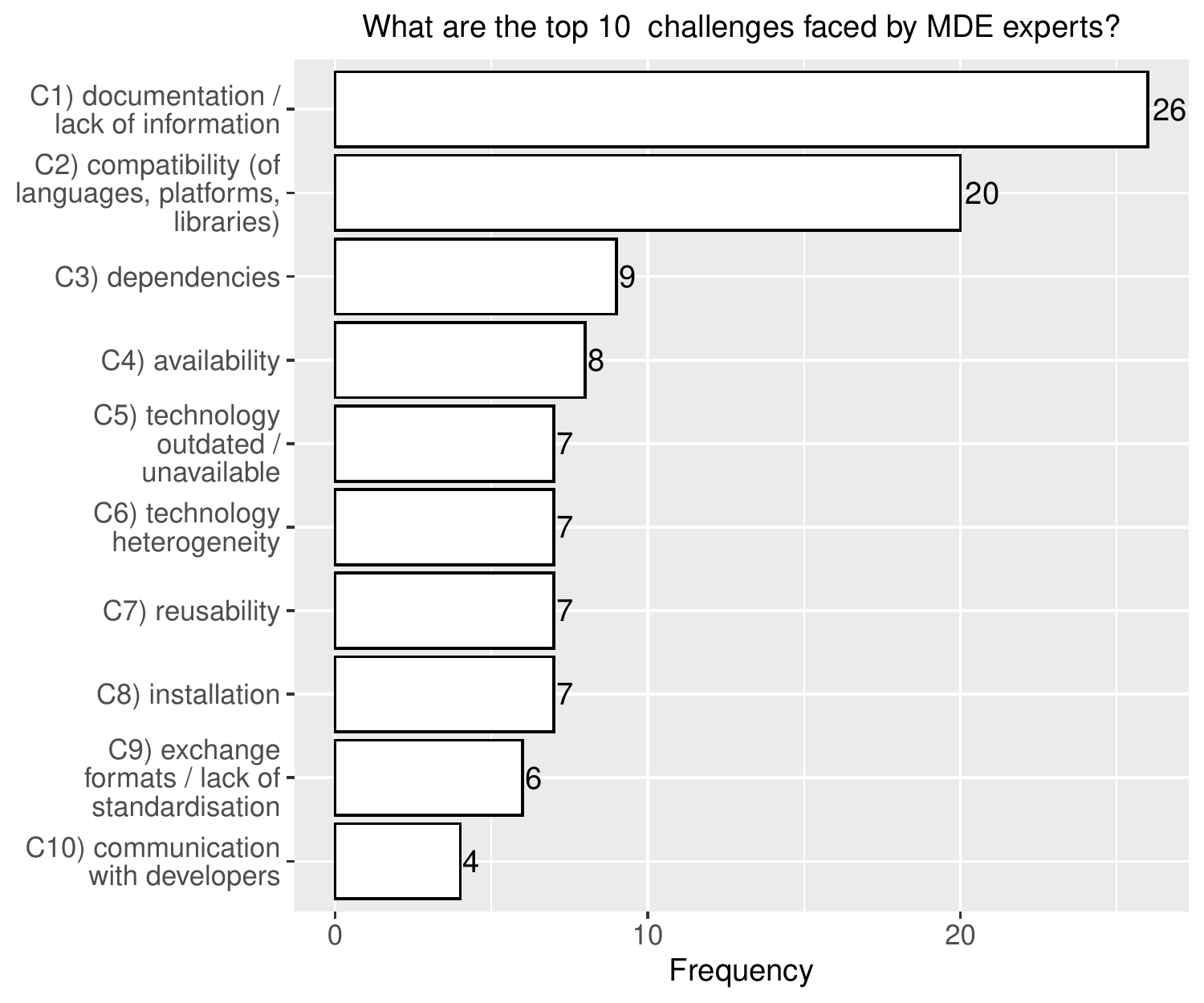}
    \caption{Top 10 challenges faced in MDE artifact sharing}
    \label{fig:top10_challenges}
\end{figure}

To analyze the challenges reported by MDE experts, we employed open coding to classify, group, and quantify answers based on their main concern. One author was responsible for the coding process; the other reviewed the assigned tags. In total, we identified 28 groups of answers from which the top ten challenges are shown in Fig.~\ref{fig:top10_challenges} with their respective identifiers.

The identified challenges largely match those identified for general SE in a previous study \cite{timperley_understanding_2020}; 
however, there are a number of noteworthy exceptions, such as \textit{technology heterogeneity}, data \textit{exchange formats}, and the \textit{lack of standardisation}, that can make MDE artifacts more difficult to be reused and produced.

Also corroborating the findings by Timperley et al. \cite{timperley_understanding_2020}, we found that \textit{the lack of information and documentation} about artifacts was the topmost challenge faced by our MDE experts. As one participant indicated:

\begin{quote} 
    \textit{``Textual description about an model can be useful to better explain the model and mitigating doubts.''}\textit{\tiny[P75]}
\end{quote}

{Human comprehension} is known to be an important task that contributes towards high-quality modeling artifacts \cite{krogstie_quality_2012}. Thus, to enhance researchers' comprehension, artifact creators should provide useful \textit{information and documentation} that describe the context of development and relevance of the artifact to the addressed problem, and facilities it offers.

In the second place, we found \textit{compatibility issues} among the topmost reported challenges. To mitigate this problem, artifact authors should always provide details about the technologies and concepts that underpin the artifact. Particularly, these should include the version identifiers of modeling languages, input file formats, or third-party artifacts used in the project, e.g., libraries, frameworks, integrated development environments.

Although reporting a detailed description of an artifact may be seen as irrelevant or not worthy \cite{timperley_understanding_2020}, finding a \textit{good-enough} amount of information can improve the quality of artifacts and facilitate their future reuse and repurposing. For example, indicating the operating system and hardware context in which the artifact was developed and tested shall support on setting up experimental environments.

% \summary{Challenges faced by MDE experts}{
% From our participants responses, we identified 28 groups of challenges.
% The utmost challenges reported was the lack of documentation and information about an artifact.
% In the second place, we found the compatibility issues between language standards, platforms, libraries.
% }

\subsubsection{To what extent do the guidelines cover them?}

In this section, we analyze the top ten challenges to identify to what extent they have been covered by our factual questions and their respective practices. In Table~\ref{tab:top10:traceability}, we depict a traceability matrix summarizing what perspectives have at least one practice able to cover a given challenge. Challenges are shown by their respective rank identifier and marked cells indicate that perspective-challenge mapping.

\begin{table}[!t]
\centering
\caption{Traceability matrix for the 5W2H perspectives and Top 10 challenges encountered by MDE experts}
\resizebox{0.49\textwidth}{!}
{%
\def\arraystretch{1.5}%  1 is the default, change whatever you need
\begin{tabular}{
% |c
|>{\centering\arraybackslash}p{0.1\linewidth}
|l|l|l|l|l|l|l|l|l|l|l|}
\hline
\multirow{2}{*}{5W2H}&
\multicolumn{1}{c|}{\multirow{2}{*}{Question}}&
\multicolumn{10}{c|}{Challenge}\\ \cline{3-12} 
&\multicolumn{1}{c|}{} 
&{C1}&{C2}&{C3}&{C4}&{C5}&{C6}&{C7}&{C8}&{C9}&{C10}\\ \hline
\multirow{3}{*}{What}&1.1) What is it all about?&{\checkmark}&&&&&&&&&\\ \cline{2-12} 
&1.2) What does it have?&{\checkmark}&&{\checkmark}&{\checkmark}&&&{\checkmark}&&&\\ \cline{2-12} 
&1.3) What underpins the artifact?&{\checkmark}&{\checkmark}&{\checkmark}&{\checkmark}&{\checkmark}&{\checkmark}&{\checkmark}&&{\checkmark}&\\ \hline
Why&2.1) Why it was created?&{\checkmark}&&&&&&&&&\\ \hline
\multirow{3}{*}{Where}&3.1) Where is it hosted?&{\checkmark}&&&{\checkmark}&&&&&&\\ \cline{2-12} 
&3.2) Where shall I cite?&{\checkmark}&&&&&&&&&\\ \cline{2-12} 
&3.3) Where to find related work?&{\checkmark}&&&&&&&&&\\ \hline
\multirow{3}{*}{Who}&4.1) Who could use it?&{\checkmark}&&&&&&{\checkmark}&&{\checkmark}&\\ \cline{2-12} 
&4.2) Who are the authors?&{\checkmark}&&&&&&&&&{\checkmark} \\ \cline{2-12} 
&4.3) Who funded this project?&{\checkmark}&&&&&&&&&\\ \hline
\multirow{2}{*}{When}&5.1) When did changes happen?&{\checkmark}&&&{\checkmark}&&&{\checkmark}&&&\\ \cline{2-12} 
&5.2) When do future changes shall happen?&{\checkmark}&&&{\checkmark}&&&{\checkmark}&&&{\checkmark} \\ \hline
\multirow{6}{*}{How}&6.1) How is it organized?&{\checkmark}&&{\checkmark}&&&&{\checkmark}&&{\checkmark}&\\ \cline{2-12} 
&6.2) How to setup a running environment?&{\checkmark}&{\checkmark}&{\checkmark}&{\checkmark}&{\checkmark}&{\checkmark}&{\checkmark}&{\checkmark}&&\\ \cline{2-12} 
&6.3) How to get started?&{\checkmark}&&&{\checkmark}&&&{\checkmark}&&&\\ \cline{2-12} 
&6.4) How to replicate the experiment?&{\checkmark}&&&{\checkmark}&&&{\checkmark}&&&\\ \cline{2-12} 
&6.5) How to run the analysis of results?&{\checkmark}&&&{\checkmark}&&&{\checkmark}&&&\\ \cline{2-12} 
&6.6) How could it be repurposed?&{\checkmark}&{\checkmark}&{\checkmark}&{\checkmark}&{\checkmark}&{\checkmark}&{\checkmark}&&&\\ \hline
How many&7.1) How many resources does it need?&{\checkmark}&&{\checkmark}&&&&{\checkmark}&{\checkmark}&&\\ \hline
\multicolumn{12}{c}{\raisebox{-.5\height} \tiny C1=Documentation ~ C2=Compatibility ~ C3=Dependencies ~ C4=Availability ~ C5=Tech. outdated/unavailable  }\\
\multicolumn{12}{c}{\raisebox{-.5\height} \tiny  C6=Tech. Heterogeneity ~ C7=Reusability ~ C8=Installation ~ C9=Exchange formats ~ C10=Communication}
\end{tabular}%
}
\label{tab:top10:traceability}
\end{table}

Guaranteeing that useful information about an artifact is available is one of the main goals of our guidelines. Hence, in all seven perspectives, we identified at least one practice that could address challenge C1. 

In the perspectives \textit{What}, \textit{How} and \textit{How Many}, we identified various practices able to address challenges from C2 to C7. Examples of these practices are 
\textit{relying on well-maintained libraries}, \textit{reporting known issues/bugs/limitations}, and \textit{indicating of library names and their respective version identifiers}. 

Artifact installation is an important step in the practices associated with questions \textit{6.2)} and \textit{7.1)}. Thus, many of their practices were found to be suitable means to cover the challenge C8, such as \textit{providing instructions to install the artifact} and \textit{indicating skills and/or settings required for artifact usage}.

To address challenge C9, we identified practices in questions \textit{1.3)}, \textit{4.1)} and \textit{6.1)}. Examples of these practices are \textit{reporting standards or specifications used to develop the artifact} and \textit{adopting open/non-proprietary files formats}.

Finally, to address challenge C10, we identified practices in questions \textit{4.2)} and \textit{5.2)}. Examples of these practices are \textit{being open for change requests and receiving feedback from users} and \textit{providing communication channels for interacting with authors and the community}.

% \summary{Challenges covered by our guidelines}{
% By analyzing our best practices, 
% we found our proposed guidelines provided reasonable means to cover the top ten challenges reported by MDE experts.
% }

\summary{Challenges (RQ2)}{
We identified 28 challenges reported by MDE experts.
The two most common challenges by far were a lack of documentation and compatibility issues between languages, platforms, and libraries.
By mapping our practices against them, we found that our guidelines provided means to reasonably cover the top 10 challenges.
}

\subsection{How do MDE experts prioritize the  practices? (RQ3)} 
\label{sec:rq3}

To evaluate the importance of our proposed guidelines, we analyzed how our participants classified the 84 practices in the three levels of priority: \textit{Essential}, \textit{Desirable}, and \textit{Unnecessary}. These alternatives were adapted from the classification schema used in the ACM SIGSOFT Empirical Standards \cite{ralph_acm_2020}.
In case of doubts or interest in omitting answers, we also provided a \textit{No answer} alternative. Based on the priorities assigned, we identified as \textit{top priority practices} all those which had at least 50\% of the participants rating it as an \textit{Essential} item.
In Table~\ref{tab:5w2h_priority}, we show the 23 top priority practices. The full list of all practices with their priorities is available in our supplementary material \cite{mdeartifacts_zenodo,mdeartifacts_github}.

We found that six out of the seven 5W2H perspectives had at least one top priority practice. We noticed that the \textit{What} and \textit{How} perspectives showed most of the top priority practices. Particularly, we found that all \textit{What} questions had at least one top priority practice. In the \textit{How} perspective, the only question that did not include any top priority practice was the ``6.6) How could it be repurposed?''.

These findings are informative for users of our guidelines, such as artifact developers and  organizers of artifact evaluation processes, as we further discuss in Sect.~\ref{sec:discussion}.

\summary{Prioritization of practices (RQ3)}{
Most participants rated most practices at least as desirable, but there is less agreement about the classification into \emph{essential vs. desirable}.
From out 84 practices, we identify a set of 23 top-priority practices that were deemed as essential by more than half of the participants.
}

\subsection{What is the quality of the proposed guidelines? (RQ4)}

\begin{figure}[!ht]
\centering
\includegraphics[width=0.94\linewidth]{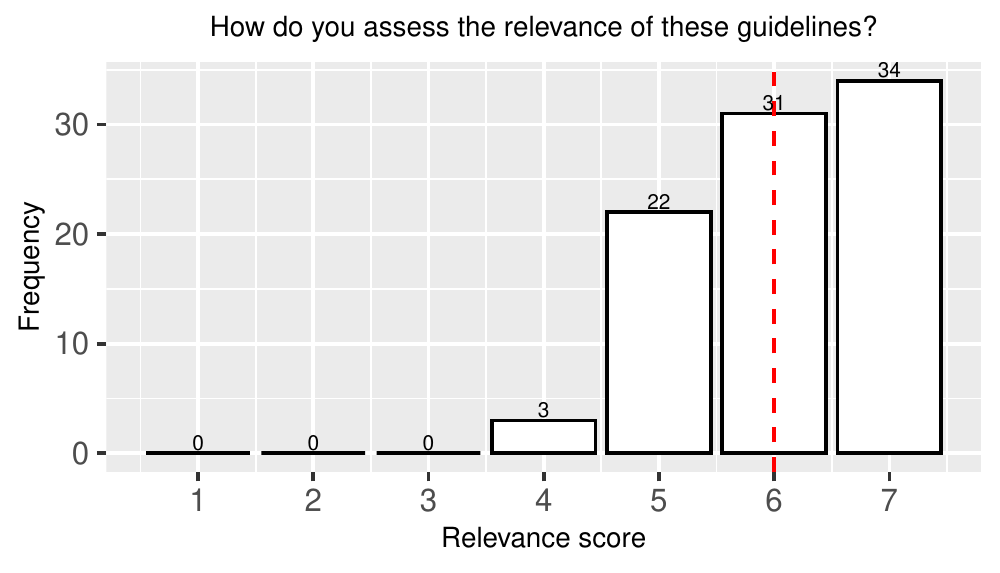}
\includegraphics[width=0.94\linewidth]{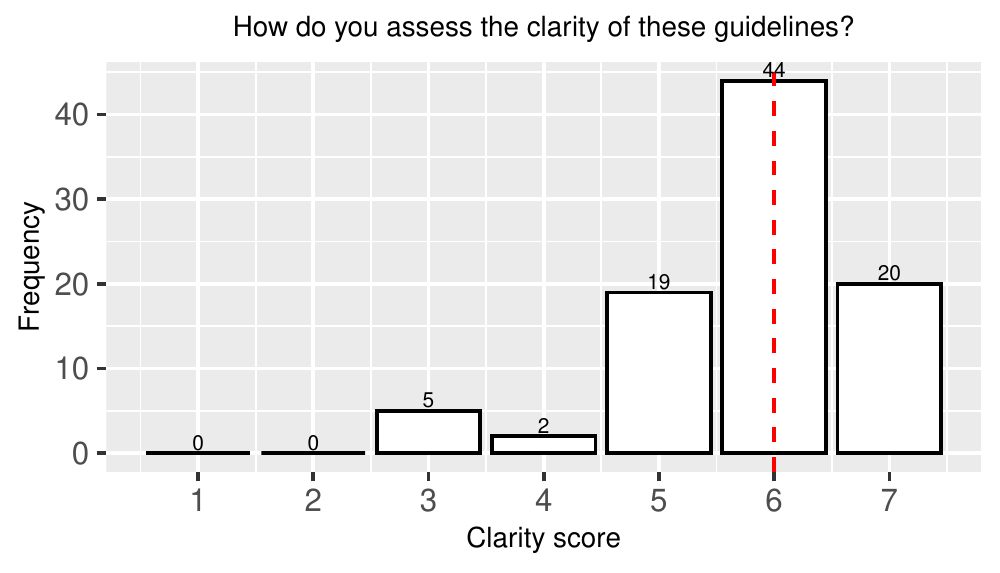}
\includegraphics[width=0.94\linewidth]{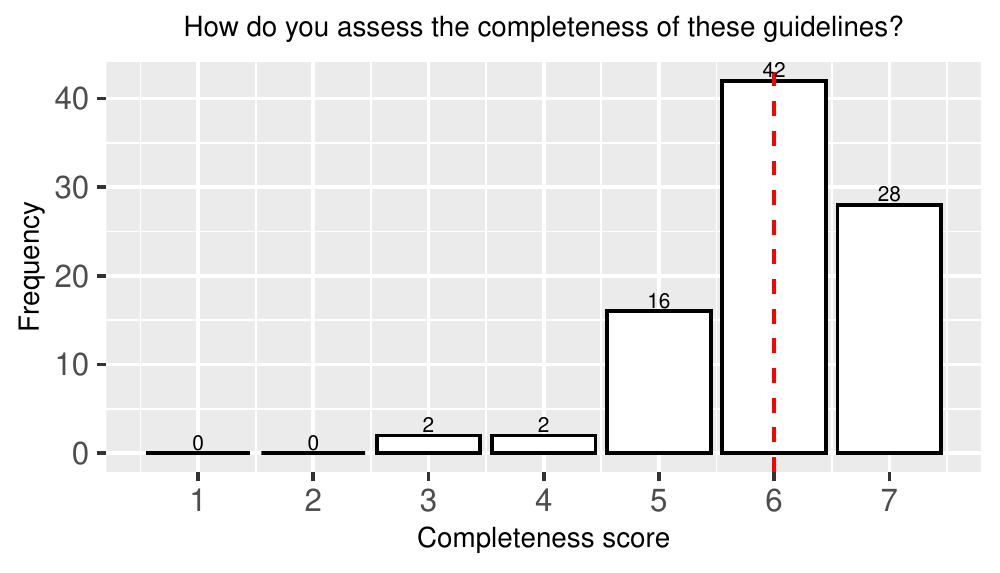}
\caption{Relevance, clarity, and completeness ratings} \label{fig:overallscores}
\end{figure}

In the last part of our questionnaire, we asked participants to provide, on a seven-points Likert scale, an overall score to the completeness, clarity, and relevance of our guidelines. In Fig.~\ref{fig:overallscores}, we show the frequency of scores for completeness, clarity, and relevance with their respective medians indicated as a vertical dashed line.

Overall, for all three dimensions, more than 92\% of our participants reported positive quality scores (5--7 in the seven-point Likert scale). For completeness, clarity, and relevance, we respectively found that 95.5\%, 92.2\%, and 96.6\% of participants reported positive scores. No participant reported a score below three points. In the textual remarks, we found positive sentiments that mirror the positive scores, for example:
\begin{quote} 
    \textit{``Broad implementation of guidelines such as these is ESSENTIAL for advancing MDE technologies and research!''}\textit{\tiny[P13]}
\end{quote}

Using the textual feedback, we analyzed the cases in which participants gave negative scores. Considering clarity, which attracted the highest number of negative scores (5.5\% of all respondents), we received comments concerning the focus of some practices on artifacts \textit{maintained for a long time} \textit{\tiny[P33]}, and the various meanings that \textit{``sharing''}\textit{\tiny[P58]} may take.

\begin{quote} 
    \textit{
		``In terms of clarity, the questions use some abstract terms, in particular sharing. I was somehow confused by this term since sharing MDE artifacts may be associated with a research paper or not. Specially when the artifact is produced in an industrial context.'' 
		%Therefore, shared artifacts can be considered with different viewpoints: providers, users, contributors, reviewers, competitors, professors, students... The last two are captured in your first questions. For the others, I would provide different answer if I were a provider, user, contributor, etc.
		}\textit{\tiny[P58]}
\end{quote}

Overall, these findings indicate that our guidelines for MDE artifact sharing were seen as reasonably complete, relevant, and clear. However, we also found there is still room for improvements, such as a need for practices considering special kinds of artifacts, stakeholders, and circumstances in which artifacts may be developed (e.g., industry, academia).

\begin{table*}[h]
    \centering
    \caption{Practices for MDE artifact sharing: 23 top-priority practices (out of 84 in total)}
    \vspace{-0.15cm}
    \resizebox{\textwidth}{!}
    {% <------ Don't forget this %
\begin{tabular}{
% >{\centering\arraybackslash}p{0.035\linewidth}
c
|l|l|c}
\hline
5W2H &
  Question &
  Practice &
  Priority \\ \hline
\multirow{6}{*}{What} &
  \multirow{3}{*}{1.1) What is it all about?} &
  Indicate the context of its development \tiny{(e.g., domain, problem, project)} &
  \hspace{-0.45cm} \raisebox{-.5\height}{\includegraphics[width=0.37\textwidth]{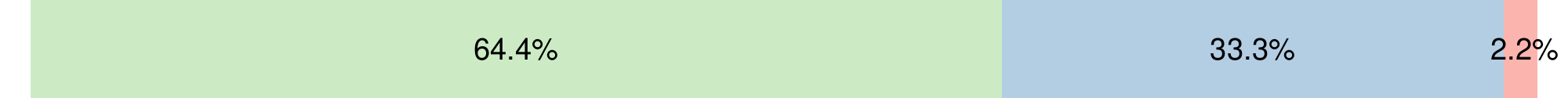}}\\ \cline{3-4} 
 &
   &
  Report its name &
  \hspace{-0.45cm} \raisebox{-.5\height}{\includegraphics[width=0.37\textwidth]{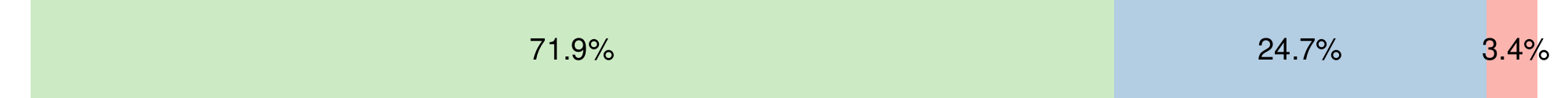}}\\ \cline{3-4} 
 &
   &
  Indicate its main functionalities supported \tiny{(e.g., modeling language, model analysis)} &
  \hspace{-0.45cm} \raisebox{-.5\height}{\includegraphics[width=0.37\textwidth]{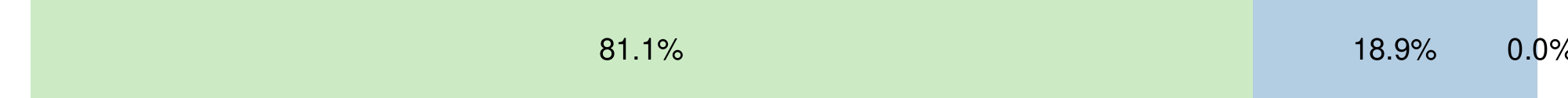}}\\ \cline{2-4} 
 &
  1.2) What does it have? &
  Include everything required for replications \tiny{(i.e., complete)} &
  \hspace{-0.45cm} \raisebox{-.5\height}{\includegraphics[width=0.37\textwidth]{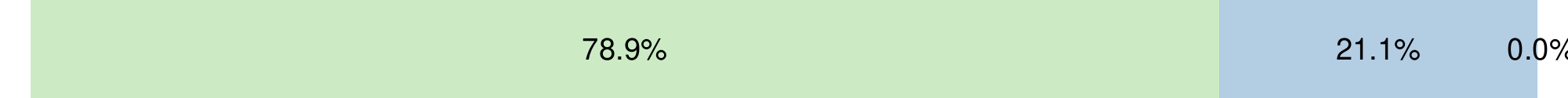}}\\ \cline{2-4} 
 &
  \multirow{2}{*}{1.3) What underpins the artifact?} &
  Indicate modeling languages used to develop it \tiny{(e.g., UML, SysML, BPMN)} &
  \hspace{-0.45cm} \raisebox{-.5\height}{\includegraphics[width=0.37\textwidth]{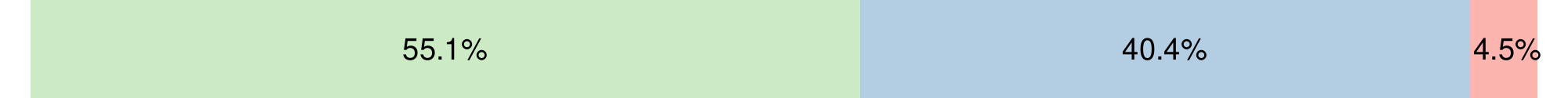}}\\ \cline{3-4} 
 &
   &
  Indicate libraries/frameworks used and their respective versions \tiny{(e.g., Eclipse release)}\hspace{-0.45cm} &
  \hspace{-0.45cm} \raisebox{-.5\height}{\includegraphics[width=0.37\textwidth]{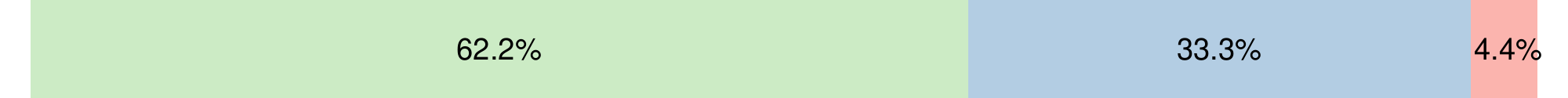}}\\ \hline
Why &
  2.1) Why it was created? &
  Indicate its objective/goal \tiny{(e.g., replicability, reusability)} &
  \hspace{-0.45cm} \raisebox{-.5\height}{\includegraphics[width=0.37\textwidth]{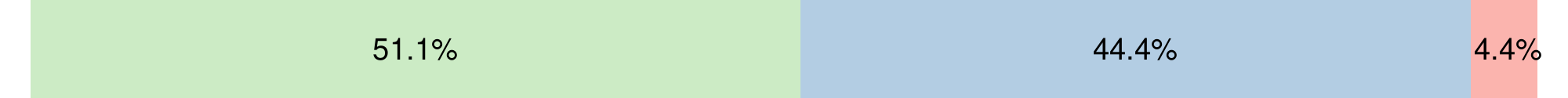}}\\ \hline
\multirow{2}{*}{Where} &
  3.1) Where is it hosted? &
  Repository is open and public \tiny{(e.g., GitHub, Zenodo, Figshare)} &
  \hspace{-0.45cm} \raisebox{-.5\height}{\includegraphics[width=0.37\textwidth]{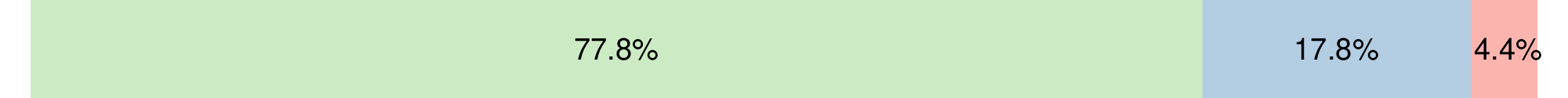}}\\ \cline{2-4} 
 &
  3.3) Where to find related work? &
  Give credit to data obtained from other sources \tiny{(e.g., author, repository)} &
  \hspace{-0.45cm} \raisebox{-.5\height}{\includegraphics[width=0.37\textwidth]{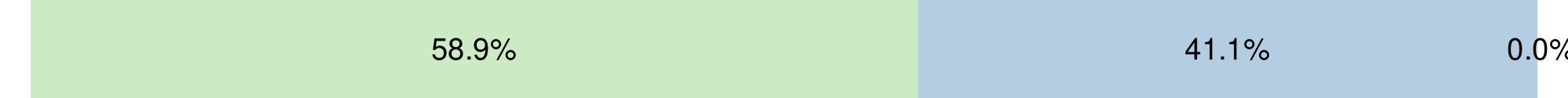}}\\ \hline
\multirow{3}{*}{Who} &
  4.1) Who could use it? &
  Deposited under an explicit open license \tiny{(e.g., reported in a LICENSE file)} &
  \hspace{-0.45cm} \raisebox{-.5\height}{\includegraphics[width=0.37\textwidth]{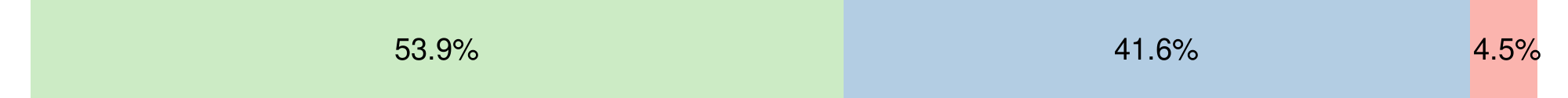}}\\ \cline{2-4} 
 &
  4.2) Who are the authors? &
  Indicate the names of its authors &
  \hspace{-0.45cm} \raisebox{-.5\height}{\includegraphics[width=0.37\textwidth]{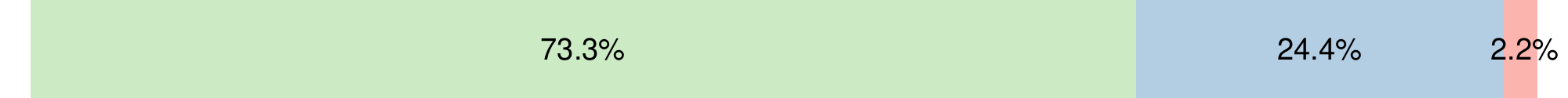}}\\ \cline{2-4} 
 &
  4.2) Who are the authors? &
  Indicate the authors's contact details \tiny{(e.g., email, ResearchGate, website)} &
  \hspace{-0.45cm} \raisebox{-.5\height}{\includegraphics[width=0.37\textwidth]{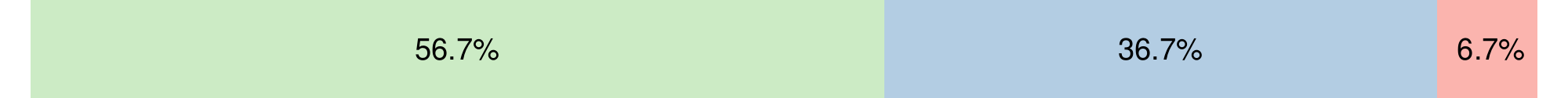}}\\ \hline
When &
  5.1) When did changes happen? &
  Tracked using version control \tiny{(e.g., GitHub, GitLab, BitBucket)} &
  \hspace{-0.45cm} \raisebox{-.5\height}{\includegraphics[width=0.37\textwidth]{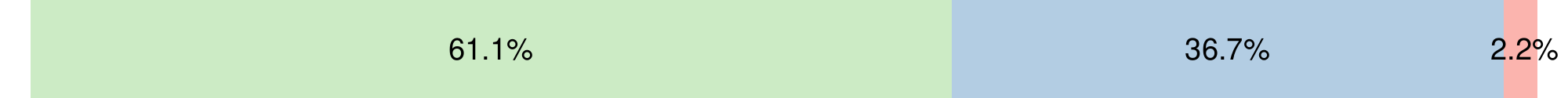}}\\ \hline
\multirow{9}{*}{How} &
  6.1) How is it organized? &
  Files and folders shall have self-explaining names matching content &
  \hspace{-0.45cm} \raisebox{-.5\height}{\includegraphics[width=0.37\textwidth]{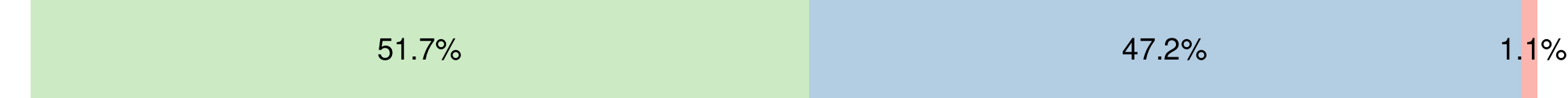}}\\ \cline{2-4} 
 &
  \multirow{3}{*}{6.2) How to setup a running environment?} &
  The artifact shall provide a step-by-step tutorial build the source code &
  \hspace{-0.45cm} \raisebox{-.5\height}{\includegraphics[width=0.37\textwidth]{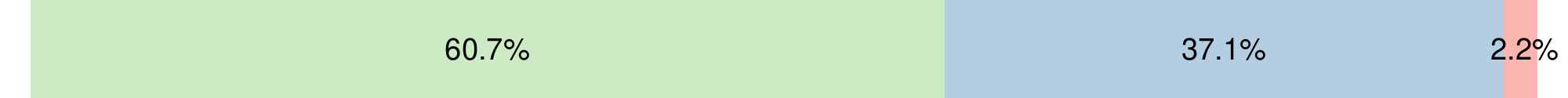}}\\ \cline{3-4} 
 &
   &
  The artifact shall provide instructions for downloading &
  \hspace{-0.45cm} \raisebox{-.5\height}{\includegraphics[width=0.37\textwidth]{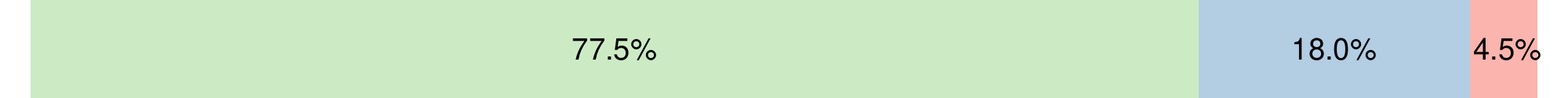}}\\ \cline{3-4} 
 &
   &
  The artifact shall provide instructions to install it &
  \hspace{-0.45cm} \raisebox{-.5\height}{\includegraphics[width=0.37\textwidth]{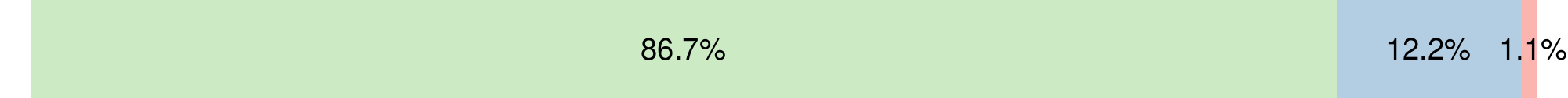}}\\ \cline{2-4} 
 &
  \multirow{2}{*}{6.3) How to get started?} &
  The artifact shall include instructions for running it on minimal test data  &
  \hspace{-0.45cm} \raisebox{-.5\height}{\includegraphics[width=0.37\textwidth]{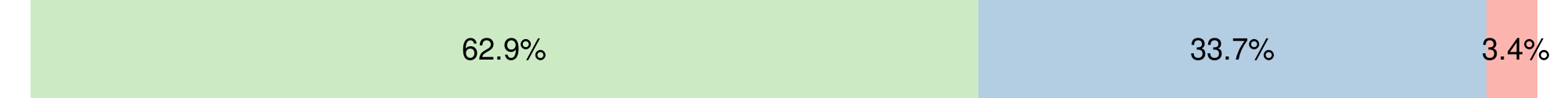}}\\ \cline{3-4} 
 &
   &
  The artifact shall include step-by-step instructions for running it \tiny{(e.g., README)} &
  \hspace{-0.45cm} \raisebox{-.5\height}{\includegraphics[width=0.37\textwidth]{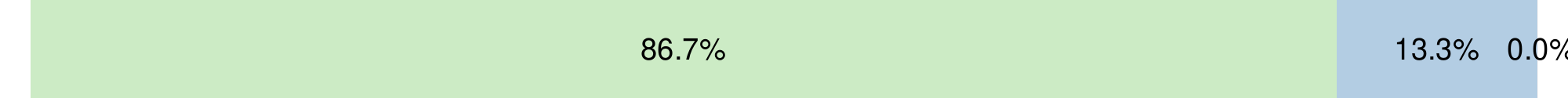}}\\ \cline{2-4} 
 &
  \multirow{2}{*}{6.4) How to replicate the experiment?} &
  Provide manual/automated instructions for the complete/partial replications &
  \hspace{-0.45cm} \raisebox{-.5\height}{\includegraphics[width=0.37\textwidth]{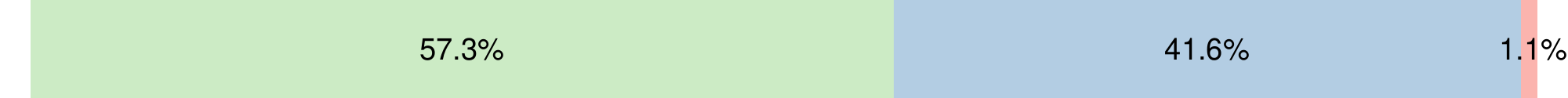}}\\ \cline{3-4} 
 &
   &
  The artifact shall include the complete set of test models considered &
  \hspace{-0.45cm} \raisebox{-.5\height}{\includegraphics[width=0.37\textwidth]{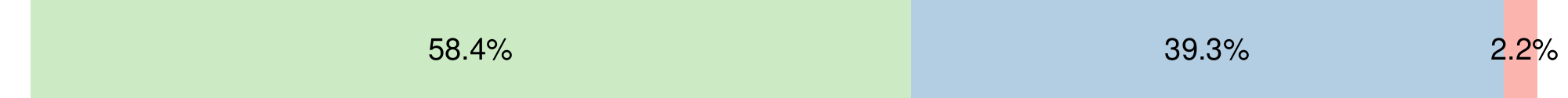}}\\ \cline{2-4} 
 &
  6.5) How to run the analysis of results? &
  Provide a clear description of measurements and metrics used in the paper &
  \hspace{-0.45cm} \raisebox{-.5\height}{\includegraphics[width=0.37\textwidth]{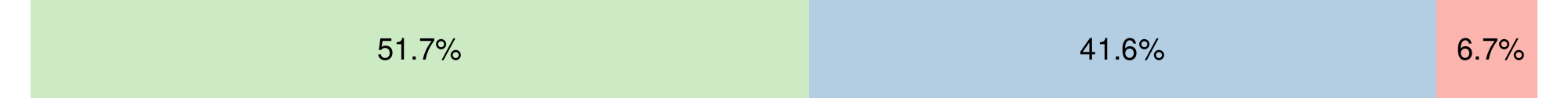}}\\ \hline
%   \begin{tabular}[c]{@{}c@{}}How\\ Many\end{tabular}
\footnotesize{How Many}
&
  7.1) How many resources does it need? &
  Indicate the system/environment settings where it was successfully evaluated &
  \hspace{-0.45cm} \raisebox{-.5\height}{\includegraphics[width=0.37\textwidth]{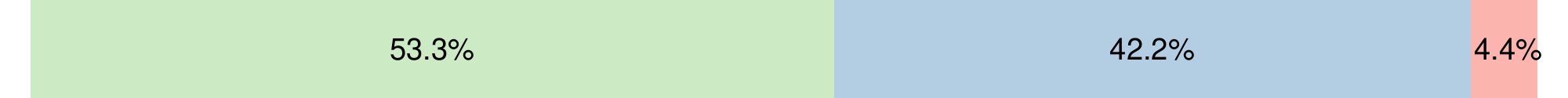}}\\ \hline
  \multicolumn{4}{r}{ Priority legend: 
  \raisebox{.5\height}{\fcolorbox{black}{lik_green}{ }} {Essential} 
  \raisebox{.5\height}{\fcolorbox{black}{lik_blue} { }} {Desirable} 
  \raisebox{.5\height}{\fcolorbox{black}{lik_red}  { }} {Unnecessary} 
  }
\end{tabular}
}
    \label{tab:5w2h_priority}
    \vspace{-0.5cm}
\end{table*}

\summary{Evaluated Quality (RQ4)}{
The surveyed MDE experts assess the completeness, relevance, and clarity of our guidelines largely positive, with between 92.2\% and 96.6\% positive scores in each dimension. We identified a number of improvement opportunities.
}

\section{Discussion}
\label{sec:discussion}

\smallskip
\noindent\textbf{Implications for artifact authors: }
Our guidelines have been designed as a toolkit to support researchers on the creation, sharing, and maintenance of artifacts in MDE research. The priority levels identified with our survey report what are the most important practices (i.e., essential) so that authors can focus on addressing them. Moreover, the top 10 issues reported by MDE experts can also indicate frequently encountered problems in MDE research and hence, drive the authors' efforts on mitigating them. 

\smallskip
\noindent\textbf{Implications for artifact evaluation organizers and reviewers:}
As previous surveys have shown \cite{hermann_community_2020,heumuller_publish_2020}, the lack of consensus on quality standards and discipline-specific guidelines for RDM and artifact sharing opens the opportunity to subjective notions of artifact quality. Our work fills this gap by complementing initiatives, such as the ACM SIGSOFT Empirical Standards, by providing MDE-specific guidelines that can also be used by artifact reviewers in MDE conferences. Since we did not find a clear agreement between participants on the prioritization, our guidelines in their current form are not intended to represent ``The'' definite list of best practices. However, they can certainly be useful to kick off the creation of venue-specific recommendations, quality criteria, or \textit{frequently asked questions} for MDE research artifacts.

% \noindent\textbf{Limitations of this study:}
% We assumed \textit{MDE artifact} as any material developed by authors and linked to a MDE research project.
% This broad definition was intentionally used to cover as many types of artifact as possible.
% As a result of this decision, at the same time that our participants indicated a reasonable satisfaction with our guidelines, 
% they also showed a need for guidelines suitable to special circumstances, such as industrial research projects, 
% where intellectual property concerns may undermine the possibility of artifact sharing and various stakeholders can be involved.
% This feedback show that the context in which an artifact is developed matters and quality guidelines should take this under consideration.

\smallskip
\noindent\textbf{Improvement opportunities:}
Based on the feedback of our participants, we found there is still room for improvement in our guidelines and noted a few possible improvements. First, employing privacy-preserving techniques, such as software obfuscation \cite{rers2019_asml_benchmark}, before sharing artifacts could be incorporated as an extraordinary industrial research practice. This could foster the disclosure of real-world artifacts in MDE industrial research. 
Second, to address the lack of viewpoint-specific practices, interviews with different stakeholders (e.g., artifact users, open-source contributors, AEC reviewers, industrial practitioners/researchers) could be done to understand which particular needs and expectations these actors may have. The interviews could provide insights for building personas and narratives about artifact development, sharing and reuse.

\subsection{Threats to validity}

We follow the recommendations by Wohlin et al. \cite{wohlin_experimentation_2012} to discuss threats to validity. Conclusion validity is out of scope, as we did not search for statistical relationships.

\smallskip
\noindent \textbf{External validity:}
These concern the generalization of our results to the overall MDE research community. The majority of our participants was formed by male academic professors. Although this may pose threats to the external validity, our findings are still relevant as professors not only have access to their own experiences but to those of their supervised team members as well. One participant explicitly reported on the experiences encountered by his students. These results can also be considered as complementary to other surveys \cite{hermann_community_2020} which focused on relatively new scientific peer reviewers, as most of our sample was formed by professors.
Another variable that may form a threat to external validity is the small female and non-binary participation. Women's participation in open science has been increasing over time \cite{murphy_women_openscience_2020} and hence may have specific influences in artifact sharing.

\smallskip
\noindent \textbf{Internal validity:}
These threats concern issues that may indicate a causal relationship when there is none. As the validity of surveys is highly dependent on its audience's representativity, we tried to cover the MDE community via two main channels: the PlanetMDE mailing list \cite{planetmde_planetmde_2021} and by approaching authors of recent SoSyM and MODELS papers. Another variable that may constitute a threat to internal validity is the social-desirability bias \cite{wikipedia_socialdesirability_2020}. To mitigate this bias, we made our questionnaire anonymous to avoid the identification of participants, following Wohlin et al.'s recommendation \cite{wohlin_experimentation_2012}. Participants could still opt-in to receive our results by providing an email address, but were informed that we would remove the e-mail addresses from our data before analyzing them.

\smallskip
\noindent \textbf{Construct validity:}
These are concerned with the ability to draw correct conclusions about the treatment and outcomes. To mitigate this kind of threat, we focused our survey on MDE experts which should be expected to understand the specificities of MDE artifacts. Moreover, to identify potential misunderstandings, we provided means to the participants to explain their choices. From our analysis, it seems that there were no major sources of misunderstanding, which is also in line with the largely positive clarity scores obtained for RQ4.

\section{Related Work}
\label{sec:relwork}

\smallskip\noindent{}\textbf{Artifact sharing:} 
Hermann et al. \cite{hermann_community_2020} surveyed AEC members of major CS conferences and found that the community expectations of artifact quality exceed the ones expressed in calls for artifacts and reviewing guidelines. Additionally, they found there is no consensus on quality thresholds for research artifacts in general. Heumüller et al. \cite{heumuller_publish_2020} analyzed ICSE papers published from the years 2007 to 2017 and identified a positive trend towards artifact availability and a small, but statistically significant, positive correlation between linking artifacts to a paper and its citations. Timperley et al. \cite{timperley_understanding_2020} reported several high-level challenges that affect the quality of artifacts and mismatched expectations between artifact creators, users and reviewers. Our paper complements this literature by introducing a domain-specific quality guideline set and investigating the opinion of domain experts about our proposed guidelines.

\smallskip\noindent{}\textbf{The 5W2H framework:}
Jia et al. \cite{jia_5w1h_2016} proposed and reported their experiences with a 5W+1H pattern to examine systematic mapping studies from a generic set of dimensions. Their pattern is proposed as a tool for investigators to define a set of systematic, generic, and complementary RQs, enabling them to kick off and expedite the mapping study process in a well-defined manner. Prana et al. \cite{prana_categorizing_2019} investigated the problem of automated classification of README file content. Using the 5W2H framework to manually annotate README file sections, the authors show that their approach can support repository owners to improve the quality of software documentation. Zhang et al. \cite{zhang_event-based_2020} proposed an approach to generate automatic summarization of scientific literature based on 5W1H event structure and trigger word templates. Compared with existing abstracts given by authors, their approach was able to provide more detailed information, in a more convenient format. In our methodology, we manually categorized research practices as answers to factual questions following the 5W2H framework. However, the process for identifying research practices could be still automated, at least partially, by means of identifying trigger word templates for research practices.

\smallskip \noindent{}\textbf{Research Data Management:}
Perrier et al. \cite{perrier_research_2017} conducted a scoping review of RDM in academic institutions and found that studies investigating processes to improve the quality of data could potentially provide tangible guidance to researchers interested in effective data reuse. Van Eeuwijk et al. \cite{van_eeuwijk_research_2021} noted that research software is a multi-faceted asset and very diverse in terms of size, complexity and format. Hence, software sustainability policies should reflect the characteristics of different software and research domains. Marjan et al. \cite{marjan_grootveld_openaire_2018} surveyed researchers from Horizon 2020 projects and identified a need for much more tailored guidance and domain-specific standards examples of data management plans. In this work, we fill these gaps by tailoring general practices for artifact sharing to the MDE research domain. The principles underpinning our investigation can be extended to other domains.

\section{Conclusion and Future Work}
\label{sec:conclusion}

Artifact sharing is known to be helpful for researchers and practitioners to build upon existing knowledge, adopt novel contributions in practice, and increase the chances of papers receiving citations. In MDE research, there is an urge for artifact sharing as the community targets a broader use of AI-based techniques, which can only become feasible if large open datasets and confidence measures about their quality are available. In this paper, we introduce a set of quality guidelines specifically tailored for MDE research artifacts. 

Based on project management principles, we designed a catalog of 84 MDE-specific research recommendations from generic practices for artifact sharing and domain-specific literature about MDE tooling issues and modeling artifact repositories. These practices are proposed as answers to factual questions that researchers can use to systematically think about concerns in MDE artifact sharing and provide directions to additional improvement inquiring. 

In a poll among 90 MDE experts, more than 92\% positively assessed the clarity, completeness, and relevance of our guidelines. Our participants reported priority levels to our practices which can guide the research decision-making during the creation, sharing, reuse, and evaluation of MDE research artifacts. The full set of generic practices, MDE-specific guidelines, and factual questions are provided as supplementary material \cite{mdeartifacts_zenodo,mdeartifacts_github,mdeartifacts_website}. 
Particularly, we highlight our project website \cite{mdeartifacts_website} available at \mbox{\small\texttt{\url{https://mdeartifacts.github.io/}}} which can be used by artifact authors, researchers, and AECs of MDE conferences and journals. 

There are several relevant directions for future work.
First, our guidelines could still be improved by indicating in which context (i.e., artifact type) and for whom (viewpoints) a given practice should be seen as essential, desirable, or unnecessary.
%Second, it would be interesting to measure the maturity of MDE research projects over time or experiment with them in an AEC. 
Second, to determine the effect of aspects such as gender and previous experiences on the prioritization of guidelines, it would be interesting to perform sub-group analysis on our data. 
Finally, our methodology focused exclusively on MDE artifacts, but it could be easily extended to other domains. Experimenting with our methodology in other domains such as software product line engineering, in which a need for consolidated community benchmarks has been expressed \cite{struber2019facing}, would be a desirable contribution.

\bibliography{paper}

\begin{thebibliography}{10}
\providecommand{\url}[1]{#1}
\csname url@samestyle\endcsname
\providecommand{\newblock}{\relax}
\providecommand{\bibinfo}[2]{#2}
\providecommand{\BIBentrySTDinterwordspacing}{\spaceskip=0pt\relax}
\providecommand{\BIBentryALTinterwordstretchfactor}{4}
\providecommand{\BIBentryALTinterwordspacing}{\spaceskip=\fontdimen2\font plus
\BIBentryALTinterwordstretchfactor\fontdimen3\font minus
  \fontdimen4\font\relax}
\providecommand{\BIBforeignlanguage}[2]{{%
\expandafter\ifx\csname l@#1\endcsname\relax
\typeout{** WARNING: IEEEtran.bst: No hyphenation pattern has been}%
\typeout{** loaded for the language `#1'. Using the pattern for}%
\typeout{** the default language instead.}%
\else
\language=\csname l@#1\endcsname
\fi
#2}}
\providecommand{\BIBdecl}{\relax}
\BIBdecl

\bibitem{pashler_editors_2012}
H.~Pashler and E.-J. Wagenmakers, ``Editors’ introduction to the special
  section on replicability in psychological science: {A} crisis of
  confidence?'' \emph{Perspectives on Psychological Science}, vol.~7, no.~6,
  pp. 528--530, 2012, place: US Publisher: Sage Publications.

\bibitem{collberg_repeatability_2016}
C.~Collberg and T.~A. Proebsting, ``Repeatability in computer systems
  research,'' \emph{Communications of the ACM}, vol.~59, no.~3, pp. 62--69,
  Feb. 2016.

\bibitem{lung_difficulty_2008}
J.~Lung, J.~Aranda, S.~M. Easterbrook, and G.~V. Wilson, ``On the difficulty of
  replicating human subjects studies in software engineering,'' in
  \emph{Proceedings of the 30th international conference on {Software}
  engineering}, ser. {ICSE} '08.\hskip 1em plus 0.5em minus 0.4em\relax New
  York, NY, USA: Association for Computing Machinery, May 2008, pp. 191--200.

\bibitem{glanz_codematch_2017}
L.~Glanz, S.~Amann, M.~Eichberg, M.~Reif, B.~Hermann, J.~Lerch, and M.~Mezini,
  ``{CodeMatch}: obfuscation won't conceal your repackaged app,'' in
  \emph{Proceedings of the 2017 11th {Joint} {Meeting} on {Foundations} of
  {Software} {Engineering}}, ser. {ESEC}/{FSE} 2017.\hskip 1em plus 0.5em minus
  0.4em\relax New York, NY, USA: Association for Computing Machinery, Aug.
  2017, pp. 638--648.

\bibitem{esecfse_call_2011}
\BIBentryALTinterwordspacing
{ESEC/FSE}, ``Call for {Artifact} {Evaluation} {\textbar} {ESEC}/{FSE} 2011,''
  2011. [Online]. Available:
  \url{http://2011.esec-fse.org/cfp-artifact-evaluation}
\BIBentrySTDinterwordspacing

\bibitem{esecfse_esecfse_2020}
\BIBentryALTinterwordspacing
------, ``{ESEC}/{FSE} 2020 - {Artifacts} - {ESEC}/{FSE} 2020,'' 2020.
  [Online]. Available:
  \url{https://2020.esec-fse.org/track/esecfse-2020-artifacts}
\BIBentrySTDinterwordspacing

\bibitem{icse_icse_2020}
\BIBentryALTinterwordspacing
ICSE, ``{ICSE} 2020 - {Artifact} {Evaluation} - {ICSE} 2020,'' 2020. [Online].
  Available:
  \url{https://2020.icse-conferences.org/track/icse-2020-Artifact-Evaluation}
\BIBentrySTDinterwordspacing

\bibitem{splc_call_2020}
\BIBentryALTinterwordspacing
{SPLC}, ``\BIBforeignlanguage{en-US}{Call for research papers – 24th {ACM}
  {International} {Systems} and {Software} {Product} {Line} {Conference}},''
  2020. [Online]. Available:
  \url{https://splc2020.net/call-for-papers/call-for-research-papers/}
\BIBentrySTDinterwordspacing

\bibitem{models_2020}
\BIBentryALTinterwordspacing
MODELS, ``{MODELS} 2020 - {Artifact} {Evaluation} - {MODELS} 2020,'' 2020.
  [Online]. Available:
  \url{https://conf.researchr.org/track/models-2020/models-2020-artifact-evaluation}
\BIBentrySTDinterwordspacing

\bibitem{ralph_acm_2020}
P.~Ralph, S.~Baltes, D.~Bianculli, Y.~Dittrich, M.~Felderer, R.~Feldt,
  A.~Filieri, C.~A. Furia, D.~Graziotin, P.~He, R.~Hoda, N.~Juristo,
  B.~Kitchenham, R.~Robbes, D.~Mendez, J.~Molleri, D.~Spinellis, M.~Staron,
  K.~Stol, D.~Tamburri, M.~Torchiano, C.~Treude, B.~Turhan, and S.~Vegas,
  ``{ACM} {SIGSOFT} {Empirical} {Standards},'' \emph{arXiv:2010.03525 [cs]},
  Oct. 2020.

\bibitem{basili_building_1999}
V.~Basili, F.~Shull, and F.~Lanubile, ``Building knowledge through families of
  experiments,'' \emph{IEEE Transactions on Software Engineering}, vol.~25,
  no.~4, pp. 456--473, Jul. 1999.

\bibitem{von_nostitz-wallwitz_towards_2018}
I.~von Nostitz-Wallwitz, J.~Krüger, and T.~Leich, ``Towards improving
  industrial adoption: the choice of programming languages and development
  environments,'' in \emph{Proceedings of the 5th {International} {Workshop} on
  {Software} {Engineering} {Research} and {Industrial} {Practice}}, ser.
  {SER}\&{IP}'18.\hskip 1em plus 0.5em minus 0.4em\relax New York, NY, USA:
  Association for Computing Machinery, May 2018, pp. 10--17.

\bibitem{colavizza_citation_2020}
G.~Colavizza, I.~Hrynaszkiewicz, I.~Staden, K.~Whitaker, and B.~McGillivray,
  ``The citation advantage of linking publications to research data,''
  \emph{PLOS ONE}, vol.~15, no.~4, p. e0230416, Apr. 2020, arXiv: 1907.02565.

\bibitem{marjan_grootveld_openaire_2018}
\BIBentryALTinterwordspacing
M.~Grootveld, E.~Leenarts, S.~Jones, E.~Hermans, and E.~Fankhauser,
  ``\BIBforeignlanguage{eng}{{OpenAIRE} and {FAIR} {Data} {Expert} {Group}
  survey about {Horizon} 2020 template for {Data} {Management} {Plans}},'' Jan.
  2018. [Online]. Available: \url{http://doi.org/10.5281/zenodo.1120245}
\BIBentrySTDinterwordspacing

\bibitem{hermann_community_2020}
B.~Hermann, S.~Winter, and J.~Siegmund, ``Community expectations for research
  artifacts and evaluation processes,'' in \emph{Proceedings of the 28th {ACM}
  {Joint} {Meeting} on {European} {Software} {Engineering} {Conference} and
  {Symposium} on the {Foundations} of {Software} {Engineering}}.\hskip 1em plus
  0.5em minus 0.4em\relax New York, NY, USA: Association for Computing
  Machinery, Nov. 2020.

\bibitem{france_models_2006}
R.~France, J.~Bieman, and B.~H. Cheng, ``Repository for model driven
  development (remodd),'' in \emph{International Conference on Model Driven
  Engineering Languages and Systems}.\hskip 1em plus 0.5em minus 0.4em\relax
  Springer, 2006, pp. 311--317.

\bibitem{babur_labeled_2019}
\BIBentryALTinterwordspacing
O.~Babur, ``A labeled ecore metamodel dataset for domain clustering,'' 2019,
  type: dataset. [Online]. Available: \url{https://zenodo.org/record/2585456}
\BIBentrySTDinterwordspacing

\bibitem{robles2017extensive}
G.~Robles, T.~Ho-Quang, R.~Hebig, M.~R. Chaudron, and M.~A. Fernandez, ``An
  extensive dataset of {UML} models in {GitHub},'' in \emph{2017 IEEE/ACM 14th
  International Conference on Mining Software Repositories (MSR)}.\hskip 1em
  plus 0.5em minus 0.4em\relax IEEE, 2017, pp. 519--522.

\bibitem{karasneh2016online}
B.~Karasneh, ``An online corpus of {UML} design models: construction and
  empirical studies,'' Ph.D. dissertation, Leiden University, 2016.

\bibitem{ATLzoo}
\BIBentryALTinterwordspacing
eclipse.org, ``{ATL Transformation Zoo}.'' [Online]. Available:
  \url{https://www.eclipse.org/atl/atlTransformations/}
\BIBentrySTDinterwordspacing

\bibitem{struber2016scalability}
D.~Str{\"u}ber, T.~Kehrer, T.~Arendt, C.~Pietsch, and D.~Reuling, ``Scalability
  of model transformations: Position paper and benchmark set,'' in
  \emph{BigMDE'16: Workshop on Scalability in Model-Driven Engineering}, 2016,
  pp. 21--30.

\bibitem{basciani2015model}
F.~Basciani, J.~Di~Rocco, D.~Di~Ruscio, L.~Iovino, and A.~Pierantonio, ``Model
  repositories: Will they become reality?'' in \emph{CloudMDE@ MoDELS}, 2015,
  pp. 37--42.

\bibitem{wilson_good_2017}
G.~Wilson, J.~Bryan, K.~Cranston, J.~Kitzes, L.~Nederbragt, and T.~K. Teal,
  ``\BIBforeignlanguage{en}{Good enough practices in scientific computing},''
  \emph{\BIBforeignlanguage{en}{PLOS Computational Biology}}, vol.~13, no.~6,
  p. e1005510, Jun. 2017, publisher: Public Library of Science.

\bibitem{wohlin_experimentation_2012}
C.~Wohlin, P.~Runeson, M.~Höst, M.~C. Ohlsson, B.~Regnell, and A.~Wessln,
  \emph{Experimentation in Software Engineering}.\hskip 1em plus 0.5em minus
  0.4em\relax Springer Publishing Company, Incorporated, 2012.

\bibitem{ru_research_2021}
R.~University, ``\BIBforeignlanguage{en}{Research {Data} management},'' 2021,
  last Modified: 2019-06-14.

\bibitem{van_eeuwijk_research_2021}
S.~van Eeuwijk, T.~Bakker, M.~Cruz, V.~Sarkol, B.~Vreede, B.~Aben, P.~Aerts,
  G.~Coen, B.~van Dijk, P.~Hinrich, L.~Karvovskaya, M.~Keijzer-de Ruijter,
  J.~Koster, J.~Maassen, M.~Roelofs, J.~Rijnders, A.~Schroten, L.~Sesink,
  C.~van~der Togt, J.~Vinju, and P.~de~Willigen,
  ``\BIBforeignlanguage{eng}{Research software sustainability in the
  {Netherlands}: {Current} practices and recommendations},'' Zenodo, Tech.
  Rep., Feb. 2021.

\bibitem{corti_managing_2019}
L.~Corti, V.~Van~den Eynden, L.~Bishop, and M.~Woollard, \emph{Managing and
  sharing research data: a guide to good practice}, 2nd~ed.\hskip 1em plus
  0.5em minus 0.4em\relax Thousand Oaks, CA: SAGE Publications, 2019.

\bibitem{brambilla_modeldriven_2012}
M.~Brambilla, J.~Cabot, and M.~Wimmer,
  \emph{\BIBforeignlanguage{eng}{Model-driven software engineering in
  practice}}, ser. Synthesis lectures on software engineering.\hskip 1em plus
  0.5em minus 0.4em\relax San Rafael, Calif.: Morgan \& Claypool, 2012, no.~1,
  oCLC: 820461802.

\bibitem{basso_revisiting_2017}
F.~P. Basso, C.~M.~L. Werner, and T.~C. Oliveira, ``Revisiting {Criteria} for
  {Description} of {MDE} {Artifacts},'' in \emph{2017 {IEEE}/{ACM} {Joint} 5th
  {International} {Workshop} on {Software} {Engineering} for
  {Systems}-of-{Systems} and 11th {Workshop} on {Distributed} {Software}
  {Development}, {Software} {Ecosystems} and {Systems}-of-{Systems} ({JSOS})},
  May 2017, pp. 27--33.

\bibitem{krogstie_quality_2012}
J.~Krogstie, ``\BIBforeignlanguage{en}{Quality of {Models}},'' in
  \emph{\BIBforeignlanguage{en}{Model-{Based} {Development} and {Evolution} of
  {Information} {Systems}: {A} {Quality} {Approach}}}, J.~Krogstie, Ed.\hskip
  1em plus 0.5em minus 0.4em\relax London: Springer, 2012, pp. 205--247.

\bibitem{steinberg_emf_book_2008}
D.~Steinberg, F.~Budinsky, M.~Paternostro, and E.~Merks, \emph{{EMF}: {Eclipse}
  {Modeling} {Framework}, 2nd {Edition}}, 2nd~ed.\hskip 1em plus 0.5em minus
  0.4em\relax Addison-Wesley Professional., Dec. 2008.

\bibitem{whittle_taxonomy_2017}
J.~Whittle, J.~Hutchinson, M.~Rouncefield, H.~Burden, and R.~Heldal,
  ``\BIBforeignlanguage{en}{A taxonomy of tool-related issues affecting the
  adoption of model-driven engineering},''
  \emph{\BIBforeignlanguage{en}{Software \& Systems Modeling}}, vol.~16, no.~2,
  pp. 313--331, May 2017.

\bibitem{katz_challenges_2016}
R.~Katz, ``\BIBforeignlanguage{en}{Challenges in {Doctoral} {Research}
  {Project} {Management}: {A} {Comparative} {Study}},''
  \emph{\BIBforeignlanguage{en}{International Journal of Doctoral Studies}},
  vol.~11, pp. 105--125, 2016.

\bibitem{pmi_guide_2017}
P.~M. Institute, Ed., \emph{A guide to the project management body of knowledge
  / {Project} {Management} {Institute}}, 6th~ed., ser. {PMBOK} guide.\hskip 1em
  plus 0.5em minus 0.4em\relax Newtown Square, PA: Project Management
  Institute, 2017.

\bibitem{tague_quality_2005}
N.~R. Tague, \emph{The quality toolbox}, 2nd~ed.\hskip 1em plus 0.5em minus
  0.4em\relax Milwaukee, Wis: ASQ Quality Press, 2005.

\bibitem{pan_framing_1993}
Z.~Pan and G.~M. Kosicki, ``Framing analysis: {An} approach to news
  discourse,'' \emph{Political Communication}, vol.~10, no.~1, pp. 55--75, Jan.
  1993.

\bibitem{hart_fiveWs_1996}
\BIBentryALTinterwordspacing
G.~Hart, ``The five {Ws}: {An} old tool for the new task of audience
  analysis,'' \emph{Technical Communication}, vol.~43, no.~2, pp. 139--145,
  1996. [Online]. Available:
  \url{http://www.geoff-hart.com/articles/1995-1998/five-w.htm}
\BIBentrySTDinterwordspacing

\bibitem{the_royal_society_openscience_2012}
\BIBentryALTinterwordspacing
{The Royal Society}, ``\BIBforeignlanguage{en-gb}{Science {As} an {Open}
  {Enterprise}},'' Royal Society, London, Tech. Rep., 2012. [Online].
  Available:
  \url{https://royalsociety.org/-/media/policy/projects/sape/2012-06-20-saoe.pdf}
\BIBentrySTDinterwordspacing

\bibitem{mdeartifacts_zenodo}
\BIBentryALTinterwordspacing
C.~D.~N. Damasceno and D.~Strüber, ``damascenodiego/mdeartifacts.github.io:
  {Artifacts} for this paper,'' Jul. 2021. [Online]. Available:
  \url{https://doi.org/10.5281/zenodo.5109401}
\BIBentrySTDinterwordspacing

\bibitem{mdeartifacts_github}
\BIBentryALTinterwordspacing
------, ``damascenodiego/mdeartifacts.github.io,'' Jul. 2021. [Online].
  Available: \url{https://github.com/damascenodiego/mdeartifacts.github.io}
\BIBentrySTDinterwordspacing

\bibitem{mdeartifacts_website}
\BIBentryALTinterwordspacing
------, ``The {MDE} {Artifacts} project,'' 2021. [Online]. Available:
  \url{https://mdeartifacts.github.io/}
\BIBentrySTDinterwordspacing

\bibitem{acm_artifact_2020}
\BIBentryALTinterwordspacing
{ACM}, ``\BIBforeignlanguage{en}{Artifact {Review} and {Badging} -
  {Current}},'' Aug. 2020. [Online]. Available:
  \url{https://www.acm.org/publications/policies/artifact-review-and-badging-current}
\BIBentrySTDinterwordspacing

\bibitem{mendez_fernandez_open_2019}
D.~Méndez~Fernández, M.~Monperrus, R.~Feldt, and T.~Zimmermann,
  ``\BIBforeignlanguage{en}{The open science initiative of the {Empirical}
  {Software} {Engineering} journal},'' \emph{\BIBforeignlanguage{en}{Empirical
  Software Engineering}}, vol.~24, no.~3, pp. 1057--1060, Jun. 2019.

\bibitem{monperrus_how_2019}
\BIBentryALTinterwordspacing
M.~Monperrus, ``\BIBforeignlanguage{en}{How to make a good open-science
  repository?}'' Dec. 2019, section: Updates in Data. [Online]. Available:
  \url{https://researchdata.springernature.com/posts/57389-how-to-make-a-good-open-science-repository}
\BIBentrySTDinterwordspacing

\bibitem{emse_emse_2021}
\BIBentryALTinterwordspacing
EMSE, ``{EMSE} {Open} {Science} {Initiative},'' Mar. 2021, original-date:
  2018-06-28T15:35:56Z. [Online]. Available:
  \url{https://github.com/emsejournal/openscience/blob/master/README.md}
\BIBentrySTDinterwordspacing

\bibitem{emse_emse_2021-1}
\BIBentryALTinterwordspacing
{EMSE}, ``\BIBforeignlanguage{en}{{EMSE} {Open} science - {Evaluation}
  {Criteria}},'' Mar. 2021. [Online]. Available:
  \url{https://github.com/emsejournal/openscience/blob/master/review-criteria.md}
\BIBentrySTDinterwordspacing

\bibitem{katz_publish_2018}
D.~S. Katz, K.~E. Niemeyer, and A.~M. Smith, ``Publish your software:
  {Introducing} the {Journal} of {Open} {Source} {Software} ({JOSS}),''
  \emph{Computing in Science Engineering}, vol.~20, no.~3, pp. 84--88, May
  2018, conference Name: Computing in Science Engineering.

\bibitem{jors_journal_2021}
\BIBentryALTinterwordspacing
JORS, ``\BIBforeignlanguage{en}{The {Journal} of {Open} {Research} {Software} -
  {Editorial} {Policies}},'' Feb. 2021. [Online]. Available:
  \url{http://openresearchsoftware.metajnl.com/about/editorialpolicies/}
\BIBentrySTDinterwordspacing

\bibitem{nasa_nasa_2021}
\BIBentryALTinterwordspacing
{NASA}, ``{NASA} {Open} {Source} {Software},'' 2021. [Online]. Available:
  \url{https://code.nasa.gov/}
\BIBentrySTDinterwordspacing

\bibitem{tacas_tacas_2019}
\BIBentryALTinterwordspacing
{TACAS}, ``{TACAS} 2019 - {ETAPS} 2019,'' 2019. [Online]. Available:
  \url{{https://conf.researchr.org/track/etaps-2019/tacas-2019-papers#Artifact-Evaluation}}
\BIBentrySTDinterwordspacing

\bibitem{cav_artifacts_2019}
\BIBentryALTinterwordspacing
CAV, ``\BIBforeignlanguage{en-US}{Artifacts {\textbar} {CAV} 2019},'' 2019.
  [Online]. Available: \url{http://i-cav.org/2019/artifacts/}
\BIBentrySTDinterwordspacing

\bibitem{timperley_understanding_2020}
C.~S. Timperley, L.~Herckis, C.~Le~Goues, and M.~Hilton, ``Understanding and
  improving artifact sharing in software engineering research,''
  \emph{Empirical Software Engineering}, vol.~26, no.~4, p.~67, 2021.

\bibitem{planetmde_planetmde_2021}
\BIBentryALTinterwordspacing
{PlanetMDE}. planetmde - model driven engineering {ANNOUNCEMENTS} - info.
  [Online]. Available:
  \url{https://listes.univ-grenoble-alpes.fr/sympa/info/planetmde}
\BIBentrySTDinterwordspacing

\bibitem{heumuller_publish_2020}
R.~Heumüller, S.~Nielebock, J.~Krüger, and F.~Ortmeier,
  ``\BIBforeignlanguage{en}{Publish or perish, but do not forget your software
  artifacts},'' \emph{\BIBforeignlanguage{en}{Empirical Software Engineering}},
  vol.~25, no.~6, pp. 4585--4616, Nov. 2020.

\bibitem{rers2019_asml_benchmark}
M.~Jasper, M.~Mues, A.~Murtovi, M.~Schl{\"u}ter, F.~Howar, B.~Steffen,
  M.~Schordan, D.~Hendriks, R.~Schiffelers, H.~Kuppens, and F.~W. Vaandrager,
  ``Rers 2019: Combining synthesis with real-world models,'' in \emph{Tools and
  Algorithms for the Construction and Analysis of Systems}, D.~Beyer,
  M.~Huisman, F.~Kordon, and B.~Steffen, Eds.\hskip 1em plus 0.5em minus
  0.4em\relax Cham: Springer International Publishing, 2019, pp. 101--115.

\bibitem{murphy_women_openscience_2020}
M.~C. Murphy, A.~F. Mejia, J.~Mejia, X.~Yan, S.~Cheryan, N.~Dasgupta,
  M.~Destin, S.~A. Fryberg, J.~A. Garcia, E.~L. Haines, J.~M. Harackiewicz,
  A.~Ledgerwood, C.~A. Moss-Racusin, L.~E. Park, S.~P. Perry, K.~A. Ratliff,
  A.~Rattan, D.~T. Sanchez, K.~Savani, D.~Sekaquaptewa, J.~L. Smith, V.~J.
  Taylor, D.~B. Thoman, D.~A. Wout, P.~L. Mabry, S.~Ressl, A.~B. Diekman, and
  F.~Pestilli, ``\BIBforeignlanguage{en}{Open science, communal culture, and
  women’s participation in the movement to improve science},''
  \emph{\BIBforeignlanguage{en}{Proceedings of the National Academy of
  Sciences}}, vol. 117, no.~39, pp. 24\,154--24\,164, Sep. 2020.

\bibitem{wikipedia_socialdesirability_2020}
\BIBentryALTinterwordspacing
Wikipedia, ``Social-desirability bias,'' 2020, page Version {ID}: 992112847.
  [Online]. Available:
  \url{https://en.wikipedia.org/w/index.php?title=Social-desirability_bias&oldid=992112847}
\BIBentrySTDinterwordspacing

\bibitem{jia_5w1h_2016}
C.~Jia, Y.~Cai, Y.~T. Yu, and T.~H. Tse, ``\BIBforeignlanguage{en}{{5W}+{1H}
  pattern: {A} perspective of systematic mapping studies and a case study on
  cloud software testing},'' \emph{\BIBforeignlanguage{en}{Journal of Systems
  and Software}}, vol. 116, pp. 206--219, Jun. 2016.

\bibitem{prana_categorizing_2019}
G.~A.~A. Prana, C.~Treude, F.~Thung, T.~Atapattu, and D.~Lo,
  ``\BIBforeignlanguage{en}{Categorizing the {Content} of {GitHub} {README}
  {Files}},'' \emph{\BIBforeignlanguage{en}{Empirical Software Engineering}},
  vol.~24, no.~3, pp. 1296--1327, Jun. 2019.

\bibitem{zhang_event-based_2020}
J.~Zhang, K.~Li, C.~Yao, and Y.~Sun, ``\BIBforeignlanguage{en}{Event-based
  summarization method for scientific literature},''
  \emph{\BIBforeignlanguage{en}{Personal and Ubiquitous Computing}}, Apr. 2020.

\bibitem{perrier_research_2017}
L.~Perrier, E.~Blondal, A.~P. Ayala, D.~Dearborn, T.~Kenny, D.~Lightfoot,
  R.~Reka, M.~Thuna, L.~Trimble, and H.~MacDonald,
  ``\BIBforeignlanguage{en}{Research data management in academic institutions:
  {A} scoping review},'' \emph{\BIBforeignlanguage{en}{PLOS ONE}}, vol.~12,
  no.~5, p. e0178261, May 2017, publisher: Public Library of Science.

\bibitem{struber2019facing}
D.~Str{\"u}ber, M.~Mukelabai, J.~Kr{\"u}ger, S.~Fischer, L.~Linsbauer,
  J.~Martinez, and T.~Berger, ``Facing the truth: benchmarking the techniques
  for the evolution of variant-rich systems,'' in \emph{SPLC'19: International
  Systems and Software Product Line Conference}.\hskip 1em plus 0.5em minus
  0.4em\relax ACM, 2019, pp. 26:1--10.

\end{thebibliography}
\bibliographystyle{IEEEtran}

\end{document}